\documentclass[]{aastex631}

\usepackage{amsmath}

\begin{document}

\title{Engine-fed Kilonovae (Mergernovae) - \uppercase\expandafter{\romannumeral2}. 
Radiation}

\correspondingauthor{Shunke Ai}
\email{shunke.ai@whu.edu.cn}
\correspondingauthor{He Gao}
\email{gaohe@bnu.edu.cn}
\correspondingauthor{Bing Zhang}
\email{bing.zhang@unlv.edu}

\author[0000-0002-9165-8312]{Shunke Ai}
\affiliation{Department of Astronomy, School of Physics and Technology, Wuhan University, Wuhan 430072, China}

\author[0000-0003-2516-6288]{He Gao}
\affiliation{Institute for Frontier in Astronomy and Astrophysics, Beijing Normal University, Beijing 102206, China}
\affiliation{Department of Astronomy, Beijing Normal University, Beijing 100875, China}

\author[0000-0002-9725-2524]{Bing Zhang}
\affiliation{Nevada Center for Astrophysics, University of Nevada Las Vegas, Las Vegas, NV 89154, USA}
\affiliation{Department of Physics and Astronomy, University of Nevada Las Vegas, Las Vegas, NV 89154, USA}







\begin{abstract}
The radioactive power generated by materials within the ejecta of a binary-neutron-star (BNS) merger powers an optical transient known as a kilonova. When the central remnant of a BNS merger is a long-lived magnetar, it continuously produces a highly magnetized wind, altering both the dynamics and temperature of the ejecta, leading to the expected emergence of an engine-fed kilonova. In the first paper of this series, we conducted a detailed study of the dynamics of wind-ejecta interaction and the efficiency of energy injection through shocks. In this work, we combine this dynamical evolution with both shock-heating and additional X-ray irradiation to model photon diffusion within a constant-opacity ejecta. By calculating the radiation, we obtain the light curve and spectral energy distribution (SED). Our findings reveal that, with energy injection, a blue bump typically appears in the early stages ($\lesssim 1$ day). Furthermore, if the magnetar has not spun down by that time, a brightening in the later stages occurs. Despite this, in a large parameter space, the expected luminosity of the engine-fed kilonova is not significantly higher than the typical r-process kilonova due to limited heating efficiency. The SED of engine-fed kilonovae peaks in the relatively blue band in the early stages and evolves towards the red, but at a slower rate compared to the typical r-process kilonova.
\end{abstract}

\keywords{Transient sources(1851) --- Magnetars(992) --- Gamma-ray bursts(629)}


\section{Introduction}
Kilonova has garnered significant attention in recent years, particularly since the detection of AT2017gfo in association with the binary-neutron-star (BNS) merger GW170817 \citep[e.g.][and the references there in, for a review]{arcavi2017,chornock2017,cowperthwaite2017,drout2017,evans2017,gao2017,kasen2017,kilpatrick2017,nicholl2017,shappee2017,smartt2017,tanvir2017,villar2017,metzger2019}. Other kilonova candidates have been observed in association with short-duration gamma-ray bursts (GRBs), which are believed to be also of a neutron star merger origin \citep[e.g.][]{tanvir13,berger13,fan13,yang15,jin16,gao2015,gao17a}. 
Most recently, some unusual GRB events, specifically GRB 211211A and GRB 230307A, have long duration but with kilonova-like emission signal associated \citep{rastinejad2022,yang2022,troja2022,gao2022,zhu2022,yangyh2023,gillanders2023,levan2023}. The James Webb Space Telescope's mid-infrared follow-up of GRB 230307A revealed an emission line at 2.15 microns in the spectrum, which is interpreted as tellurium \citep{levan2023,gillanders2023}. Additionally, the rapidly decaying bolometric luminosity of the kilonova at the later stage supports the presence of lanthanide in the ejecta. This kilonova signal is hence generally believed to be powered by r-process radioactive heating \citep{yangyh2023}.

Notably, the presence of a plateau in the X-ray afterglow light curve of GRB 230307A, detected by the Lobster Eye Imager for Astronomy (LEIA), a pathfinder of the Einstein Probe mission, indicates that the merger remnant is a magnetar \citep{sun2023,ling2023}. Previous studies have speculated that a post-merger magnetar could drive an engine-fed kilonova (mergernova), which is anticipated to be significantly brighter than a typical r-process kilonova \citep{yu2013,metzgerpiro2014,gao2015,ren2019,sarin2022,wang2023}. Therefore, the detection of the engine-fed kilonova signal serves as another crucial piece of evidence regarding the production of a long-lived magnetar. In reality, by adjusting the free parameters in the theoretical model of an engine-fed kilonova, the predicted luminosity of thermal radiation from the merger ejecta can span several orders of magnitude. When assuming a relatively low spin-down luminosity and low heating efficiency, the peak luminosity of the engine-fed kilonova can be reduced to approximately $10^{42}~{\rm erg/s}$ \citep[e.g.][for the kilonovae associated with GW170817 and GRB 211211A]{lisz2018,yu2018,ren2019,yang2022}, which can also be achieved within the framework of a typical r-process kilonova, given a reasonably large ejecta mass. Hence, for certain critical scenarios, it becomes extremely challenging to discern scenarios with or without energy injection. In some previous studies, a pulsar wind nebula (PWN) has been introduced by considering the reverse shock (RS) propagating in a magnetar wind dominated by electron-positron pairs \citep{metzgerpiro2014,ren2019,murase2018,wang2023}. They have comprehensively modeled the generation of non-thermal radiations from the PWN and how these non-thermal radiations are absorbed by the ejecta. In their models, the PWN does not significantly alter the dynamics of the merger ejecta, thus the peak time of the kilonova might remain roughly unchanged. A considerable portion of the non-thermal photons would escape from the ejecta, serving as compelling evidence for the presence of a post-merger magnetar \citep{murase2018}.

We have made continuous efforts to more precisely model the engine-fed kilonova. Our first paper in this series \citep{ai2022} provided a detailed study for the dynamics of the ejecta resulting from the interaction between the magnetar wind and the merger ejecta. It is under the framework of the mechanical model for magnetized relativistic blastwaves \citep{ai2021}. For convenience, we henceforth refer to \cite{ai2022} as Paper I. In Paper I, we found that a significant fraction ($\xi$ as high as $\sim 0.6$) of the energy from the post-merger magnetar's wind can be converted into the kinetic energy of the ejecta's bulk motion. Meanwhile, the forward shock excited in the ejecta provides an additional heating source during its propagation, with a heating efficiency of $\xi_t \sim (0.006 - 0.3)$. In this work, building upon the dynamical model proposed in Paper I, we focus on radiation. The basic formula for modeling both the heating and radiation processes of the engine-fed kilonova is presented in Section \ref{sec:methods}. The results, including the multi-band light curves and the spectral energy distribution (SED), are shown in Section \ref{sec:results}. The conclusion and discussion are detailed in Section \ref{sec:conclusion}.

\section{Methods}
\label{sec:methods}

The interaction of magnetar wind and the merger ejecta excites a forward shock (FS) in the ejecta and a reverse shock (RS) in the highly magnetized wind. A blastwave typically refers to the region situated between the FS and RS. The magnetization parameter ($\sigma_w$) of the wind likely remains significantly greater than unity all the way into the magnetar magnetosphere. The RS would propagate rapidly through the wind and quickly vanish near the magnetosphere, where the pressures of the unshocked wind and the blastwave become comparable \citep{zhangkobayashi05,mazhang2022}. We set this location as the inner boundary of for our dynamics calculation and consider the region extending from this inner boundary to the FS as the blastwave. In Paper I, we studied in detail the dynamics of this system and the heating of the ejecta due to the FS. In that paper, the unshocked ejecta was assumed to be wind-like, with a density profile of $\rho_{\rm ej} \propto r^{-2}$, where $r$ represents the distance from the central engine in the lab frame. We also assumed that the ejecta is expanding like a shell, where the speed along radial direction is constant throughout the space. Additionally, since radioactive heating and radiation do not significantly influence the dynamics, we neglected both of them, as well as photon diffusion inside the ejecta. Therefore, the evolution of the internal energy of the blastwave (equivalent to the evolution of integrated gas pressure $P_{\rm sph}$ as described in Paper I) depends solely on shock heating and the adiabatic expansion of the ejecta.

In this paper, to model the radiation of an engine-fed kilonova, we must comprehensively consider the evolution of the ejecta's internal energy. Generally, this evolution can be expressed as \citep{yu2013}.\footnote{For intuitive purposes, we directly present the evolution of internal energy here, rather than the integrated gas pressure $P_{\rm sph}$. The equation describing the evolution of $P_{\rm sph}$ should be modified accordingly, as demonstrated in Section \ref{sec:shock-heating}, where the mechanical model for blastwaves is introduced.}
\begin{eqnarray}
    \frac{d E_{\rm int}^{\prime}}{dt^{\prime}} = L_{\rm ra}^{\prime} + L_{\rm ext}^{\prime} - L_{e}^{\prime} - p\frac{dV^{\prime}}{dt^{\prime}},
\end{eqnarray}
where $L_{\rm ra}^{\prime}$ represents the the radioactive heating from the r-process nucleosynthesis. $L_{\rm ext}^{\prime}$ represents the contribution of additional heating sources beyond radioactive heating. In our work, this term specifically refers to the heating powered by the post-merger NS (magnetar). This contribution consists of two components: (1) heating from a forward shock induced in the ejecta by the highly magnetized NS wind, and (2) heating from X-rays generated from the NS wind, either through the self-dissipation of the magnetic field or from a nebula powered by the reverse shock excited there. These two additional heating components are detailed in Sections \ref{sec:shock-heating} and \ref{sec:X-ray irradiation}, respectively. $L_e^{\prime}$ represents the bolometric luminosity of kilonova radiation. $V^{\prime}$ is the volume of the ejecta. The superscript `${\prime}$' means the physical quantity is defined in the fluid's rest frame and hereafter. 

For the convenience of numerical calculation, we discretize the ejecta into $N$ shells and number the shells from $n = 1$ to $n = N$. An arbitrary physical quantity $Q$ for a certain shell $n$ is denoted as $\delta Q_n$. Sometimes we drop the subscript $n$ when referring to any shell in general. For each shell, the evolution of internal energy follows 
\begin{eqnarray}
\frac{d(\delta E_{\rm int}^{\prime})}{dt^{\prime}} = \delta L_{\rm ra}^{\prime} + \delta L_{\rm ext}^{\prime} - \delta L_{\rm out}^{\prime} - \delta L_e^{\prime} - p \frac{d(\delta V^{\prime})}{dt^{\prime}} 
\label{eq:Eint_delta}
\end{eqnarray}
where $\delta L_{\rm out}^{\prime}$ represents the net outflow of the internal energy from a certain shell due to photon diffusion. The details for photon diffusion across different shells will be discussed in Section \ref{sec:photon_diffusion}.

Another difference with Paper I is that here we introduce a gradient for bulk-motion velocity of the unshocked ejecta, to more realistically describe the outflow properties. Suppose every shell of the merger ejecta initially has a specific bulk motion velocity $v$,  in the lab frame. The velocity difference between two adjacent shells is denoted as $\delta v$. We assume the density of the ejecta depends on the velocity, which is given by
\begin{eqnarray}
\rho^{\prime} (v) \propto v^{-\alpha}.
\end{eqnarray}
Also assuming the ejecta is isotropic, the volume of a shell can be calculated as
\begin{eqnarray}
    \delta V^{\prime} &=& \Gamma \delta V \nonumber \\ &=& 4\pi R^2 \Gamma \delta v t \nonumber \\
    &=& 4\pi \Gamma v^2 \delta v t^3
\end{eqnarray}
where $R = vt$ is the distance from the central engine to the considered shell and $\Gamma$ is the Lorentz factor of the shell's bulk motion.  $t$ is the time since the shell was ejected in the central engine frame. The mass in this shell can be written as
\begin{eqnarray}
\delta M^{\prime}_{{\rm ej}} &=& \rho^{\prime} \delta V^{\prime} \nonumber \\
&\approx& 4\pi K v^{-\alpha + 2} t^3 \delta v,
\end{eqnarray}
with a normalization constant
\begin{eqnarray}
K = \left \{ \begin{array}{ll} 
\frac{(-\alpha+3)M_{\rm ej}^{\prime}}{4\pi (v_{\rm max}^{-\alpha+3} - v_{\rm min}^{-\alpha+3})t^3}, & ~~\textrm{for $ \alpha \neq 3$},\\
\\
\frac{M_{\rm ej}^{\prime}}{4\pi {\rm ln}(v_{\rm max}/v_{\rm min}) t^3}, & ~~\textrm{for $ \alpha = 3$}.
\\
\end{array} \right.
\end{eqnarray}
where $M_{\rm ej}^{\prime} = \sum \delta M^{\prime}_{{\rm ej}}$ is the total ejecta mass. Here, $\Gamma \approx 1$ is assumed for the non-relativistic initial bulk-motion velocity. Next, we discuss each term on the right-hand side of Equation \ref{eq:Eint_delta} separately.

\subsection{Radioactive heating}
The radioactive power released through nuclear reaction has been studied in detail in the literature \cite[e.g.][]{lipaczyski98,metzger2010, korobkin2012,rosswog2018,zhu2021,wu2022,chen2024}. Here we adopt the analytical treatment proposed by \cite{korobkin2012}. For unit mass, the power can be described as a function of time, which reads
\begin{eqnarray}
\dot{\epsilon}(t^{\prime}) = \epsilon_0 \left(\frac{1}{2}-\frac{1}{\pi} {\rm arctan}\frac{t^{\prime}-t_0}{\sigma}\right)^{\beta},
\end{eqnarray}
where $\epsilon_0 = 4\times 10^{18}{\rm erg~g^{-1}~s^{-1}}$, $t_0 = 1.3s$, $\sigma = 0.11s$ and $\beta = 1.3$. The power is carried by neutrinos, gamma-ray photons and mildly relativistic electrons, where only a fraction can be converted to thermal energy \citep[e.g.][]{barnes16,hotokezaka2016,kasen2019,ricigliano2024}, so that a thermalization efficiency $\epsilon_{\rm th}$ is defined. Hence, the radioactive heating power for a certain ejecta shell is expressed as 
\begin{eqnarray}
    \delta L_{\rm ra}^{\prime} = \delta M_{\rm ej}^{\prime} \dot{\epsilon}(t^{\prime}) \epsilon_{\rm th}(t^{\prime},x).
\end{eqnarray}
Here we use an analytical formula to estimate $\epsilon_{\rm th}$ as \citep{barnes16,ricigliano2024}
\begin{eqnarray}
    \epsilon_{\rm th}(t,x) = 0.36\left[{\rm exp}(-aX) + \frac{{\rm ln}(1+2bX^d)}{2bX^d}\right],
\end{eqnarray}
where $X = t(1-x^2)^{-1}$ and $x =(R - R_0) / (R_{\rm max} - R_{\rm min})$ is the position of the shell in the ejecta. The values of parameters $a$ and $b$ can be found from \cite{barnes16}, depending on the mass and velocity of the ejecta. In this paper, the ejecta mass is adopted as $M_{\rm ej} = 10^{-2}M_{\odot}$, while the maximum and minimum initial velocity of the ejecta are adopted as $v_{\rm max} = 0.2c$ and $v_{\rm min} = 0.05c$, where the mean velocity is $~0.1c$, which corresponds to $a = 1.43$, $b = 0.17$, $d = 1.46$. 

\subsection{Shock heating}
\label{sec:shock-heating}
The interaction between the post-merger NS wind and the merger ejecta can excite an FS in the ejcta, which can both alter the dynamics of the ejecta and provide an additional heating source. The shock heating rate is denoted as $L_{\rm sh}^{\prime}$ and should be included into the extra heating term $L_{\rm ext}^{\prime}$. The key to studying shock heating lies in solving the dynamics of the blastwave, as detailed in Paper I.

In Paper I, we assumed that the bulk motion velocity and the magnetization parameter for the unshocked NS (magnetar) wind are constant in both time and space, as they are believed not to significantly influence the energy injection process, especially when they were sufficiently large. The luminosity of the unshocked NS wind was also assumed to be constant as our primary focus was on examining the dependence of energy injection efficiency on the wind's luminosity. However, in this paper, while maintaining the same setups for the bulk motion and magnetization of the unshocked wind, we introduce a more realistic approach where the luminosity evolves in accordance with the spindown of the central NS, which is given by
\begin{eqnarray}
    L_w (t) = \eta L_{\rm sd,0} \left(1 + \frac{t}{t_{\rm sd}}\right)^{-2},
\end{eqnarray}
where $t_{\rm sd} = E_{\rm tot} / L_{\rm sd,0}$ is the spin-down timescale of the NS due to magnetic dipole radiation, with $E_{\rm tot}$ being the total energy budget of the NS wind and $L_{\rm sd,0}$ being the initial spin-down luminosity. We assume that the wind is generated and accelerated near the light cylinder of the newly-born millisecond magnetar at $r_{\rm acc} = 10^7 {\rm cm}$. Beyond this radius, both the Lorentz factor for the bulk motion of the NS wind ($\Gamma_w$) and its magnetization parameter ($\sigma_w$) are considered constant in both space and time for the unshocked wind. As has been elaborated in Paper I, the results remain insensitive to the exact value of $\Gamma_w$ and $\sigma_w$ when they are sufficiently large, and thus $\Gamma_w = 10^3$ and $\sigma_w = 10$ are adopted in this study. A beaming factor $\eta$ is introduced, as the fast rotation of the central NS can focus roughly $90\%$ of the spin-down power into the direction of the spin axis \citep{wang2024}. Therefore, $\eta = 0.1$ is adopted here in the scenario with nearly isotropic merger ejecta. The physical quantities preceding the RS, i.e. the wind density and magnetic strength, can be derived directly from the wind luminosity at a given time and the position of the RS ($r_r$) (See Equations 31 and 32 in our Paper I). 
The RS in the wind reaches the magnetosphere of the magnetar quickly after it was exicited and subsequently dissipates there (set as inner boundary). Then $\Gamma_{w,{\rm acc}} = \Gamma$ should be satisfied, where $\Gamma_{w,{\rm acc}}$ and $\Gamma$ represent the Lorentz factor for the bulk motion of the unshocked wind at $r_{\rm acc}$ and the bulk motion of the blastwave, respectively. With $r_r = r_{\rm acc}$ and $\Gamma_w = \Gamma$, and assuming a constant $\sigma_w$,  the physical quantities of the wind at the inner boundary, where the wind is generated, can be calculated at each time.


To study the blastwave system, one should, in principle, set the initial point of contact between the magnetar wind and the merger ejecta. In Paper I, we designated the initial position of the contact discontinuity between the wind and ejecta as $r_0$. Physically speaking, this is a catch-up problem between the wind and the ejecta, as the ejecta is launched during the merger while the continuous wind is generated after the central rigidly rotating NS has stabilized. It is believed that a post-merger NS will experience the differentially rotating phase within at least several hundred millisecond after the merger \citep[e.g.][]{radice2018}.
Therefore, here we assume a time delay between the ejecta launching and the generation of the constant NS wind ($\delta t_l$; in the central engine frame). Then, we can calculate the time of the first contact between the NS wind and the ejecta, which reads as
\begin{eqnarray}
t_c \approx \delta t_l + \frac{\delta t_l \beta_{\rm min}}{1-\beta_{\rm min}} = \frac{\delta t_l}{1-\beta_{\rm min}},
\end{eqnarray}
The corresponding first-contact position should be
\begin{eqnarray}
r_0 \approx \frac{\delta t_l}{(1-\beta_{\rm min})} \beta_{\rm min} c.
\end{eqnarray}
When $\beta_{\rm min} = 0.05$, $t_c = 1.05 (\delta t_l / 1{\rm s})~{\rm s}$ and $r_0 = 1.57 \times 10^9(\delta t_l / 1{\rm s})~{\rm cm}$. Although the exact value of $t_l$ is unknown, since the magnetar wind lasts for a duration much longer than $t_l$, the value of $t_l$ is unlikely to have a substantial influence on the results. We tested with $t_l = 0.1$s, $1$s, and $10$s, respectively, and found little difference in the outcomes. Consequently, we use $t_l = 1$s consistently throughout the paper for simplicity.

For the unshocked ejecta, we introduced a gradient for the initial bulk-motion velocity, as discussed in the beginning of Section \ref{sec:methods}, and assume that the bulk-motion velocity of each shell remains constant over time. The velocity of the material immediately preceding the FS can be simply calculated as $v_f = r_f/t$, where $r_f$ is the radius of the FS and $t$ is time in the lab frame. The density upstream of the FS should be\footnote{To maintain consistency with Paper I, the symbol ``$\rho$'' will henceforth represent the density in the rest frame of the local fluid, without the need for a superscript ``$\prime$''.}
\begin{eqnarray}
\rho_1 = K v_f^{-\alpha} = K (r_f/t)^{-\alpha},
\end{eqnarray}
and the gas pressure is
\begin{eqnarray}
p_1 = \frac{\delta E_{\rm int}^{\prime}(v_f, t)}{ 3 \delta V^{\prime} (v_f,t)}.
\end{eqnarray}
The internal energy density in the unshocked ejecta is much smaller than the rest mass energy density. Therefore, the jump conditions for cold upstreams can still be applied for the FS. The gradient of velocity for the materials behind the shock is assumed to vanish.

In Paper I, we defined the integrated pressure over the entire blastwave region, assuming a spherical geometry, as $P_{\rm sph} = \Gamma P_{\rm shp}^{\prime} = \int_{r_r}^{r_f} 4\pi r^2 p dr$, where $r_r$ and $r_f$ represent the position of the RS (inner boundary) and FS, respectively. $\Gamma$ is the bulk motion Lorentz factor of the blastwave. The evolution of the integrated pressure is crucial for calculating the dynamics of the wind-ejecta interaction. Generally, considering all the effects discussed in this paper, it should be modified as
\begin{eqnarray}
\frac{dP_{\rm sph}^{\prime}}{dr_d} &=& \frac{dP_{\rm sph}^{\prime}}{dr_d}\bigg|_{\rm rs} + \frac{dP_{\rm sph}^{\prime}}{dr_d}\bigg|_{\rm fs} + \frac{dP_{\rm sph}^{\prime}}{dr_d}\bigg|_{\rm exp} \nonumber \\ 
&&+ \frac{dP_{\rm sph}^{\prime}}{dr_d}\bigg|_{\rm ra} + 
\frac{dP_{\rm sph}^{\prime}}{dr_d}\bigg|_e + \frac{dP_{\rm sph}^{\prime}}{dr_d}\bigg|_{\rm diff},
\label{eq:dPshp}
\end{eqnarray}
where $dr_d = \beta dt$ represents the evolution of the position of the contact discontinuity, and $\beta$ is the dimensionless velocity of the blastwave's bulk motion. The first three terms on the right-hand side of Equation \ref{eq:dPshp} represent the accumulation of $P_{\rm sph}$ due to the reverse and forward shocks, as well as the effect of adiabatic expansion, respectively. These topics have been discussed in detail in Paper I. Here, we clarify the physical meaning of the last three terms and derive their expressions.

The contribution of radioactive heating to the blastwave can be further divided into two parts: from the already swept materials and the newly swept shell, respectively. This is expressed as
\begin{eqnarray}
\frac{dP_{\rm sph}^{\prime}}{dr_d}\bigg|_{\rm ra} = \frac{\hat{\gamma}-1}{\Gamma \beta c} L_{\rm ra,s}^{\prime} + 4\pi r_f^2 \frac{\beta_f - \beta}{\beta}p_1,
\end{eqnarray}
where
\begin{eqnarray}
L_{\rm ra,s}^{\prime} = M_{{\rm ej},s}^{\prime}\dot{\epsilon}(t^{\prime})\epsilon_{\rm th}(t^{\prime},x)
\end{eqnarray}
with the swept mass as
\begin{eqnarray}
M_{{\rm ej},s}^{\prime} = \int_{v_{\rm min}}^{v_f} K v^{-\alpha} dv.
\end{eqnarray}
Here $v_f$ ($\beta_f$) represents the dimensionless speed for the propagation of the forward shock, and $\hat{\gamma} = 4/3$ is adopted as the adiabatic index for the matter in the blastwave. Similarly, the term related to radiation is expressed as
\begin{eqnarray}
\frac{dP_{\rm sph}^{\prime}}{dr_d}\bigg|_e &=& - \frac{\hat{\gamma}-1}{\Gamma \beta c} \int_r^{r_f} \frac{\partial L_e^{\prime}}{\partial r}dr \nonumber \\
&=& -\frac{\hat{\gamma}-1}{\Gamma \beta c} \sum_{n=1}^{N} \delta L_{e,n}^{\prime}
\end{eqnarray}
with discretization. The loss of integrated pressure due to the photon diffusion across the FS is written as 
\begin{eqnarray}
    \frac{dP_{\rm sph}^{\prime}}{dr_d}\bigg|_{\rm diff} = - \frac{\hat{\gamma}-1}{\Gamma \beta c} L_{\rm diff, n\rightarrow n+1}^{\prime} \bigg|_{v_n = v_f}.
\end{eqnarray}
The luminosities for photon diffusion across different layers ($L_{\rm diff}^{\prime}$) and of thermal radiation ($\delta L_e^{\prime}$) are detailed studied in Section \ref{sec:photon_diffusion} and Section \ref{sec:radiation}, respectively.

Although the profiles of the physical quantities in the ejecta have been neglected when calculating the dynamics, they are not neglected when calculating the radiation. The evolution of internal energy still follows Equation \ref{eq:Eint_delta}. The shock heating luminosity could be expressed as
\begin{eqnarray}
    L_{\rm sh}^{\prime} = 4\pi r_f^2 (\beta_f - \beta) c \Gamma^2 \frac{p_f}{\hat{\gamma}-1} \xi_{\rm sh}
\label{eq:Lsh}
\end{eqnarray}
where $p_f$ can be obtained by solving the jump conditions of the FS. 
Here we introduce a factor $\xi_{\rm sh}$ to describe the internal energy generated by the shock that can be converted to thermal photons, which could be estimated as
\begin{eqnarray}
    \xi_{\rm sh} = 1 - {\rm exp}\left( - \tau\right),
\end{eqnarray}
where $\tau$ is the optical depth for photons at the shock crossing place to diffuse out of the ejecta. The calculation of $\tau$ will be presented in Section \ref{sec:radiation}. Shock heating only happens in location in the shell where the FS is crossing through (denoted as the $k$th shell), therefore 
\begin{eqnarray}
    \delta L_{\rm sh}^{\prime} = \delta_n^k L_{\rm sh}^{\prime}
\end{eqnarray}
where 
\begin{eqnarray}
\delta_n^k = \left \{ \begin{array}{ll} 1, & ~\textrm{for $ k = n$},\\
\\
0, & ~\textrm{for $ k \neq n$}.
\end{array} \right.
\end{eqnarray}

Under the mechanical model for blastwaves, the distance from the ejecta's $n$th shell to the central engine is written as
\begin{eqnarray}
R_n =\left \{ \begin{array}{ll} 
v_n t, & ~~\textrm{for $ v_n > v_s (t)$},\\
\\
(R_{n,s} - R_{n,d}) + r_d, & ~~\textrm{for $v_n < v_s (t)$},\\
\end{array} \right.
\end{eqnarray}
$R_{n,s}$ and $R_{n,d}$ represent the radius of shell $n$ and the radius of the contact discontinuity, respectively, at the time when shell $n$ was shocked. $r_d$ represents the radius of the contact discontinuity at time $t$. The velocity $v_s$ as a function of $t$ represents the initial bulk motion velocity of the shell being shocked by the FS, and it can be calculated using the formula $v_s = r_f/t$. The volume of the shell in the rest frame can be directly expressed as
\begin{eqnarray}
\delta V_n^{\prime} = 4 \pi R_n^2 \Delta_n^{\prime}
\end{eqnarray}
with the thickness of the shell
\begin{eqnarray}
\Delta_n^{\prime} =\left \{ \begin{array}{ll} 
\Gamma_{\rm ej,n} t \delta v_n, & ~~\textrm{for $ v_n > v_s(t)$},\\
\\
\Gamma t_{n,s} \delta v_n, & ~~\textrm{for $v_n < v_s(t)$},\\
\end{array} \right.
\end{eqnarray}
where $t_{n,s} = R_{n,s}/v_n$ is the time when shell $n$ was swept by the FS. The evolution of the volume of the shell could be expressed as
\begin{eqnarray}
\frac{d(\delta V_n^{\prime})}{dt^{\prime}} =\left \{ \begin{array}{ll} 
12\pi \Gamma_{\rm ej,n}^2 v_n^2 t^2 \delta v_n, & ~~\textrm{for $ v_n > v_s(t)$},\\
\\
8\pi \Gamma R_n \Delta_n^{\prime} v_n + 4 \pi R_n^2 \Delta_n^{\prime} \frac{d\Gamma}{dt}, & ~~\textrm{for $v_n < v_s(t)$}.\\
\end{array} \right.
\label{eq:V_evolution}
\end{eqnarray}
The pressure in each shell is thus $p = (1/3)(\delta E_{\rm int}^{\prime} / \delta V^{\prime})$.

\subsection{X-ray irradiation}
\label{sec:X-ray irradiation}
In Paper I, we assumed that the RS crosses the mangetar's wind and vanishes within a very short timescale due to the wind's high magnetization, preventing the formation of a long-lasting pulsar wind nebula. However, some authors argued that the magnetar's wind is not purely magnetic flux, but also contains a large fraction of electron-positron pairs. The synchrotron radiation emitted by the RS-accelerated pairs could convert the wind's magnetic energy into X-ray photons \citep{metzgerpiro2014,wang2023}. Furthermore, the self-dissipation of the magnetic field may be another channel for producing X-ray photons \citep{zhangyan2011}. Regardless, these X-ray photons would be absorbed by the merger ejecta and contribute to the kilonova radiation. In this study, although the NS wind is assumed to be dominated by the magnetic field, we do not completely ignore the potential X-ray irradiation. We introduce another efficiency parameter, $\eta_X$, to represent the fraction of wind luminosity that can be converted into X-ray radiation. The X-ray heating rate, denoted as $L_X^{\prime}$, should also be included in the extra heating term $L_{\rm ext}^{\prime}$.

Assume that the X-ray heating occurs only in the first $k_{1,f}$ shells from the inner to the outer ejecta, and distributed uniformly as 
\begin{eqnarray}
    \delta L_{X,t,n}^{\prime} = \left \{ \begin{array}{ll} 
L_{X,t}^{\prime} \frac{\delta V_n^{\prime}}{\sum_{i=1}^{k_{1,f}} \delta V_n^{\prime}} , & ~~\textrm{for $ n \leq k_{1,f}$},\\
\\
0, & ~~\textrm{for $n > k_{1,f}$},\\
\end{array} \right.
\end{eqnarray}
where $L_{X,t}^{\prime} = \xi_X \eta_X L_w / \Gamma^2$. The subscript ``$1$''and ``$f$'' in $k_{1,f}$ indicate that the X-ray photons propagate from the innermost shell (the $1$st shell) in the 'forward' direction. $k_{1,f}$ is obtained by modeling the photon diffusion, by equaling the timescale for the photons diffusing from the innermost layer to the $k_{1,f}$th layer, with the time resolution for our calculation. This will be detailed presented in Section \ref{sec:photon_diffusion}. Similar as Equation \ref{eq:Lsh}, a thermalization factor $\xi_X$ is introduced. Again, it is expressed as
\begin{eqnarray}
    \xi_X = 1 - {\rm exp}\left(-\tau_{\rm ej}\right),
\end{eqnarray}
where $\tau_{\rm ej}$ is the optical depth for photons diffusing from the innermost to the outermost layers. 

\subsection{Photon diffusion}
\label{sec:photon_diffusion}
The diffusion of photons across various layers in radiation-dominated materials can be described using the heat conduction equation \citep{mihalas1984}. Ideally, one should concurrently solve the equations governing the dynamics and photon diffusion.
However, the timescale for photons to effectively diffuse between shells depends on the evolving optical depth, which is not always commensurate with the time resolution used to calculate the dynamics. The timescale for photons to diffuse across a width, characterized by an initial velocity gradient $\delta v$, can be approximately estimated as $t_{{\rm diff},\delta v}^{\prime} \sim (l/c)(\delta \tau)^2$ when $\delta \tau > 1$, where $\delta \tau$ represents the optical depth corresponding to $\delta v$, and $l = 1/(\kappa\rho^{\prime})$ is the mean free path of the photons. Given that $\delta \tau = \kappa\rho^{\prime} t \delta v$ and $\rho^{\prime} \propto v^{-\alpha} t^{-3}$, we can derive that $t_{{\rm diff},\delta v}^{\prime} \propto v^{-\alpha} t^{-1}$. It is clear that the diffusion timescale for outer layers is significantly shorter than that for inner layers, and both decrease as the merger ejecta expands. In other words, the diffusion rate for the inner layers is higher, and this rate increases as the ejecta expands.  Consequently, the time step for numerically solving the heat conduction equation should be adjusted at each instant. In this study, rather than numerically solving the heat conduction equation, we employ a simplified approach to approximate photon diffusion.

Suppose the ejecta is divided into $N$ shells where the $N$th shell is the outermost. Under the Lagrangian description, 
the diffusion time of photons across layers from the $i$th to the $j$th can be estimated as
\begin{eqnarray}
    t_{\rm diff, i\rightarrow j}^{\prime} = \frac{l}{c}{\cal N}_{\rm s,i \rightarrow j},
\end{eqnarray}
where ${\cal N}_s$ is the expected number of scatterings (absorptions) for a photon to propagate through these shells. For the $n$th shell, within a time step $\Delta t^{\prime}$, the photon can diffuse across $k_{n,f}$ shells forwards and $k_{n,b}$ shells backwards. The expected number of scattering for a photon to propagate across the shells can be estimated as
\begin{eqnarray}
    {\cal N}_{s,n \leftrightarrow n+k_{n,f}} = {\rm max}\left\{ \sum_{i=n}^{n+k_{n,f}} \delta \tau_i, ~\left(\sum_{i=n}^{n+k_{n,f}} \delta \tau_i^2\right)^2 \right\}
\end{eqnarray}
and ${\cal N}_{s,n-k_{n,f} \leftrightarrow n}$ as the same, where $\delta \tau = \kappa \rho^{\prime} \delta v t$. $\kappa$ represents the opacity. $k_{n,f}$ and $k_{n,b}$ can be obtained by equaling $\Delta t^{\prime}$ with $t_{\rm diff}$. Suppose at each time step $\Delta t^{\prime}$, from the $(n - k_{n,b})$th to $(n + k_{n,f})$th shells, local thermal equilibrium tends to be achieved. Therefore, the luminosity of photons diffusing across the interface between the $n$th and $(n+1)$th shells is calculated as 
\begin{flalign}
    L_{\rm diff,n\rightarrow n+1}^{\prime} = \frac{\sum_{i = n-k_{n,b}}^{n}\delta E_{\rm int,i}^{\prime} - \sum_{i = n-k_{n,b}}^{n + k_{n,f}+1}\delta E_{\rm int,i}^{\prime}\frac{\sum_{i = n-k_{n,b}}^{n}\delta V_{i}^{\prime}}{\sum_{i = n-k_{n,b}}^{n + k_{n,f}+1}\delta V_{i}^{\prime}}}{\Delta t^{\prime}}.    
\end{flalign}
The photon diffusion term in Equation \ref{eq:Eint_delta} can be written as
\begin{eqnarray}
    \delta L_{\rm out,n}^{\prime} = L_{\rm diff, n\rightarrow n+1}^{\prime} -  L_{\rm diff, n-1 \rightarrow n}^{\prime}
\end{eqnarray}

When calculating photon diffusion, we only focus on the shocked and unshocked ejecta, as the other system components contribute little. Consequently, there exist two boundaries for the ejecta: the innermost and outermost layers. Although photons are able to diffuse across different shells of the ejecta, for the kilonova photons, the innermost and the outermost boundaries are assumed to be ``walls." In other words, we have
\begin{eqnarray}
    \delta L^{\prime}_{\rm out,N} = -  L_{\rm diff, N-1 \rightarrow N}^{\prime}
\end{eqnarray}
and
\begin{eqnarray}
    \delta L^{\prime}_{\rm out,1} = L_{\rm diff, 1 \rightarrow 2}^{\prime}.
\end{eqnarray}
The leakage of photons from the boundary is effectively addressed by incorporating the radiation term ($\delta L_e^{\prime}$) into Equation \ref{eq:Eint_delta}. Note that these two boundaries are ``free" for X-ray photons, allowing them to pass freely through the ejecta's boundaries.

\subsection{Radiation}
\label{sec:radiation}
Given that the ejecta has been divided into $N$ shells. The optical depth for photons in the $n$th shell to diffuse out of the ejecta is 
\begin{eqnarray}
    \tau_n = \sum_n^N \delta \tau_n,
\end{eqnarray}
where a fraction of photons ($e^{-\tau_n}$) can freely diffuse out of the ejecta without any scattering (absorption). In the time step $dt^{\prime}$, the photons are expected to be scattered for ${\cal N}_{s,r} = c dt / l$ times, so that the photons which can escape from the ejecta only experience less than ${\cal N}_{s,r}$ scatters. Therefore, at each time step, the fraction of energy that would be released is estimated as 
\begin{eqnarray}
    f_{\rm rel} = 1 - (1 - {\rm exp}\left(-\tau\right))^{{\cal N}_s}.
\end{eqnarray}
The contribution of the $n$th shell to the total bolometric luminosity then reads as
\begin{eqnarray}
    \delta L_{e,n}^{\prime} = \frac{\delta E_{\rm int,n}^{\prime} f_{\rm rel,n}}{dt^{\prime}}, 
\end{eqnarray}
while the total is calculated as 
\begin{eqnarray}
    L_e = \sum_{n = 0}^{N} {\cal D}_n^2 \delta L_{e,n}^{\prime},
\end{eqnarray}
where ${\cal D}_n = 1/\left[\Gamma_n(1-\beta_n)\right]$ is the Doppler factor.
Assume the radiation from different shells all have black-body spectra, but with different effective temperature, which reads as
\begin{eqnarray}
    T^{\prime} = \left( \frac{\delta E_{\rm int}^{\prime}}{a \delta V^{\prime}} \right)^{1/4}.
\end{eqnarray}
In the observer's frame, the luminosity in one specific frequency can be expressed as
\begin{eqnarray}
    \nu L_{\nu} = \sum_{n=1}^{N} {\cal K}_n \frac{(h \nu / {\cal D}_n)^4}{ {\rm exp}(h \nu / {\cal D}kT_{\rm n}^{\prime})-1}
\end{eqnarray}
where the normalization coefficient reads as 
\begin{eqnarray}
    {\cal K}_n = \frac{{\cal D}_n^2\delta L_{e,n}^{\prime}}{\int \frac{(h \nu / {\cal D}_n)^4}{ {\rm exp}(h \nu / {\cal D}kT_{\rm n}^{\prime})-1} d\nu}
\end{eqnarray}
Considering all heating, radiation and photon diffusion terms introduced above, the overall evolution of the ejecta's internal energy could be calculated following Equation \ref{eq:Eint_delta}. The evolution of the volume of each shell is described with Equation \ref{eq:V_evolution}. In the ejecta's comoving frame, lab frame and observer's frame, the times intervals' relation can be written as $dt^{\prime} = dt / \Gamma = {\cal D}_N dt_{\rm obs}$.

\subsection{Numerical resolution}

To simulate photon diffusion and radiation processes, we have segmented the ejecta into $N$ shells, representing the spatial resolution. Fundamentally, spatial and temporal resolutions are independent parameters. However, we found that having the FS traverse one shell at each time step simplifies the calculations. Thus, in this work, the spatial and temporal resolutions are correlated.

Given the initial selection of the number of shells ($N$) for the ejecta, before the breakout of the FS, the time resolution ($\Delta t$) needs to be updated at each instant. Specifically, it is determined as $\Delta t = \Delta_s / (v_f - v_s)$, where $\Delta_s = t \delta v_s$ represents the shell thickness that the FS is traversing in the lab frame. After the FS breaks out of the ejecta, we fix $\Delta t$ to the value it had when the FS traversed the outermost shell. It is worth noting that this fixed time resolution is defined on a logarithmic scale.

To achieve a precise light curve, adopting an appropriate time resolution is crucial. However, manually setting $\Delta t$ is challenging, as even with a fixed $N$, the time resolution varies across different parameter sets for the magnetar wind and the merger ejecta. Although we cannot formulate an analytical relation between spatial and temporal resolution, their positive correlation is undisputed. Initially, we choose $N = 1000$ shells for each parameter set and follow the aforementioned model to calculate the luminosity evolution. The calculations begin at $t_c$ and terminate at $t_{\rm end} = 10^7~{\rm s}$, with the timeline divided into roughly uniform intervals on a logarithmic scale. We set a minimum of $N_t = 300$ time intervals as our criterion. If the results yield a lower time resolution, we increase the spatial resolution until the time resolution criteria is met. Furthermore, we test convergence by increasing $N_t$ to $500$, but find no significant difference in the light curve compared to when $N_t = 300$ is used.

\section{Results}
\label{sec:results}
The outcome of a typical r-process kilonova depends on the ejecta's properties, including the total mass ($M_{\rm ej}$), opacity ($\kappa$) and velocity gradient. In the engine-fed scenario, in addition to these ejecta properties, those related to the central engine would be even more crucial, i.e. the total energy budget of magnetar's wind ($E_{\rm tot}$) and the initial spin-down luminosity ($L_{\rm sd,0}$). The efficiency ($\eta_X$) with which the spin-down luminosity is converted into X-ray irradiation is another crucial parameter. For this study, we have fixed the parameters pertaining to the ejecta and are solely examining the impact of energy injection. 

The rotational kinetic energy of the newly born magnetar can be estimated as $E_{\rm rot} = \frac{1}{2} I \Omega_0^2$, where $I$ and $\Omega_0$ represent its momentum of inertia and initial rotational angular velocity. Adopting the fiducial values that $I = 1.5\times 10^{45}~{\rm g~cm^2}$ and $\Omega_0 = 2\pi/P_0 = 6.3 \times 10^{3}~{\rm rad/s}$ with $P_0 = 1~{\rm ms}$ as the initial spin period, $E_{\rm rot}$ should be on order of $10^{52}~{\rm erg}$. Due to the possible release of energy through secular gravitational waves, the total energy budget carried by the magnetar wind ($E_{\rm tot}$) could be smaller than the rotational kinetic energy ($E_{\rm rot}$). Considering the beaming effect of the magnetar wind, the ``effective" total wind energy could be even smaller, denoted as $E_{\rm eff} = \eta E_{\rm tot}$. Therefore, we choose several values of $E_{\rm tot}$ that are equal to or less than $10^{52}~{\rm erg}$ for our discussion. Consequently, the corresponding $E_{\rm eff}$ is less than $10^{51}~{\rm erg}$, as $\eta = 0.1$ is adopted throughout this paper.

The initial spin-down luminosity ($L_{\rm sd,0}$) and the efficiency of the wind to dissipate into X-rays ($\eta_X$) are not completely free parameters. In principle, they can be inferred through the observation of X-ray plateaus on the afterglow light curve of the short GRBs \citep{rowlinson2010,rowlinson2013,lu2015}. According to the short GRB sample with an X-ray internal plateau collected in \cite{lu2015}, the break times of the plateaus concentrate around several hundred seconds, indicating that the spin-down luminosity should be $L_{\rm sd,0}\sim {\rm a~few} \times 10^{49}~{\rm erg/s}$ if the total rotational kinetic energy is $E_{\rm rot} \sim 10^{52}~{\rm erg}$. The observed X-ray plateau luminosity in the \cite{lu2015} sample  range from approximately $10^{46}~{\rm erg/s}$ to $10^{49}~{\rm erg/s}$. Incorporating a beaming factor of $\eta = 0.1$ (because the effective X-ray luminosity along the short GRB direction could be enhanced by a factoe of $\eta^{-1}=10$), the range for $\eta_X$ should be from $10^{-4}$ to ${\rm a~few} \times 10^{-2}$, with the majority of the sources having $\eta_X \sim 10^{-2}$. For $\eta_X \ll 1$, the dissipation of magnetic field would not influence the dynamics of the system, so that we will not modify the formulae related to the mechanical model of the blastwave. When $\eta_X \sim 10^{-2}$, the contribution of X-ray irradiation would be comparable with the shock heating, because the shock heating efficiency obtained from Paper I is also $\xi_t \sim 0.01$. In this work, we present the cases with $\eta_X = 10^{-2}$ and $10^{-4}$ separately. Recent observations also discovered some X-ray internal plateaus sources that are not associated with short GRBs, likely due to a relatively large observing angle (as expected theoretically \citep{zhang13}). These plateaus have an even lower inferred luminosities \citep{xue2019,sun2019,ai2021Xray}. This indicates that either a smaller $L_{\rm sd,0}$ or a smaller $\eta_X$ is allowed. For GW170817, if a long-lived post-merger NS is produced, the required spin-down luminosity should be even much smaller \citep{ai18}.

\subsection{Light curve}
\label{sec:luminosity}
The spin-down process of the post-merger magnetar is complicated, making the light curves of engine-fed kilonovae diverse. Generally, they could be divided into two cases:

\begin{itemize}

    \item Case I: The spin-down luminosity is relatively large. An early blue bump in the light curve would be expected. The luminosity at a later stage is determined by total energy injection budget and the X-ray heating efficiency $\eta_X$.

    \item Case II: The spin-down luminosity is relatively low. The early blue bump in the light curve is not that obvious, but a late-time brightening is expected.
    
\end{itemize}

\begin{figure*}
\centering
\begin{tabular}{ccc}
    \resizebox{55mm}{!}{\includegraphics[]{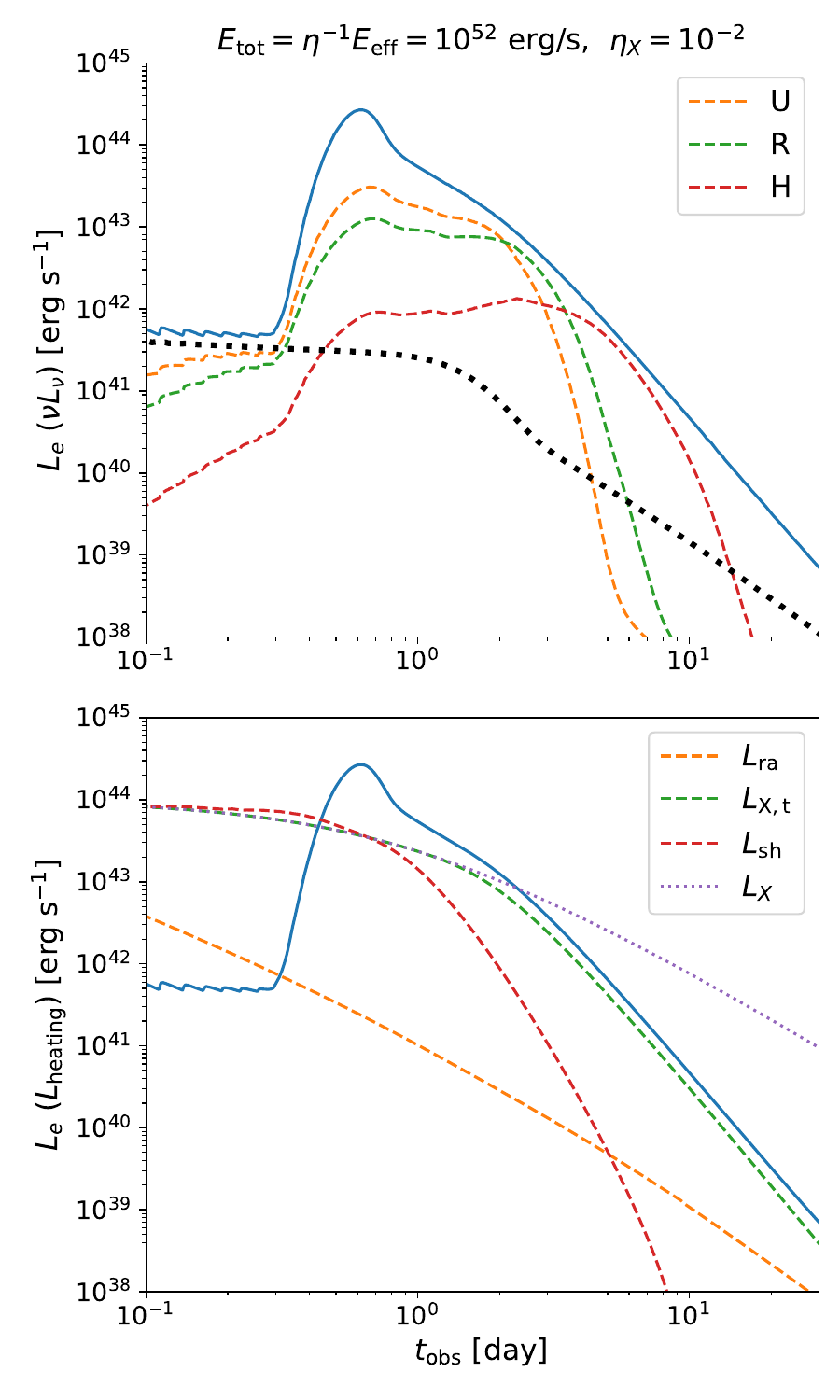}}   &  \resizebox{55mm}{!}{\includegraphics[]{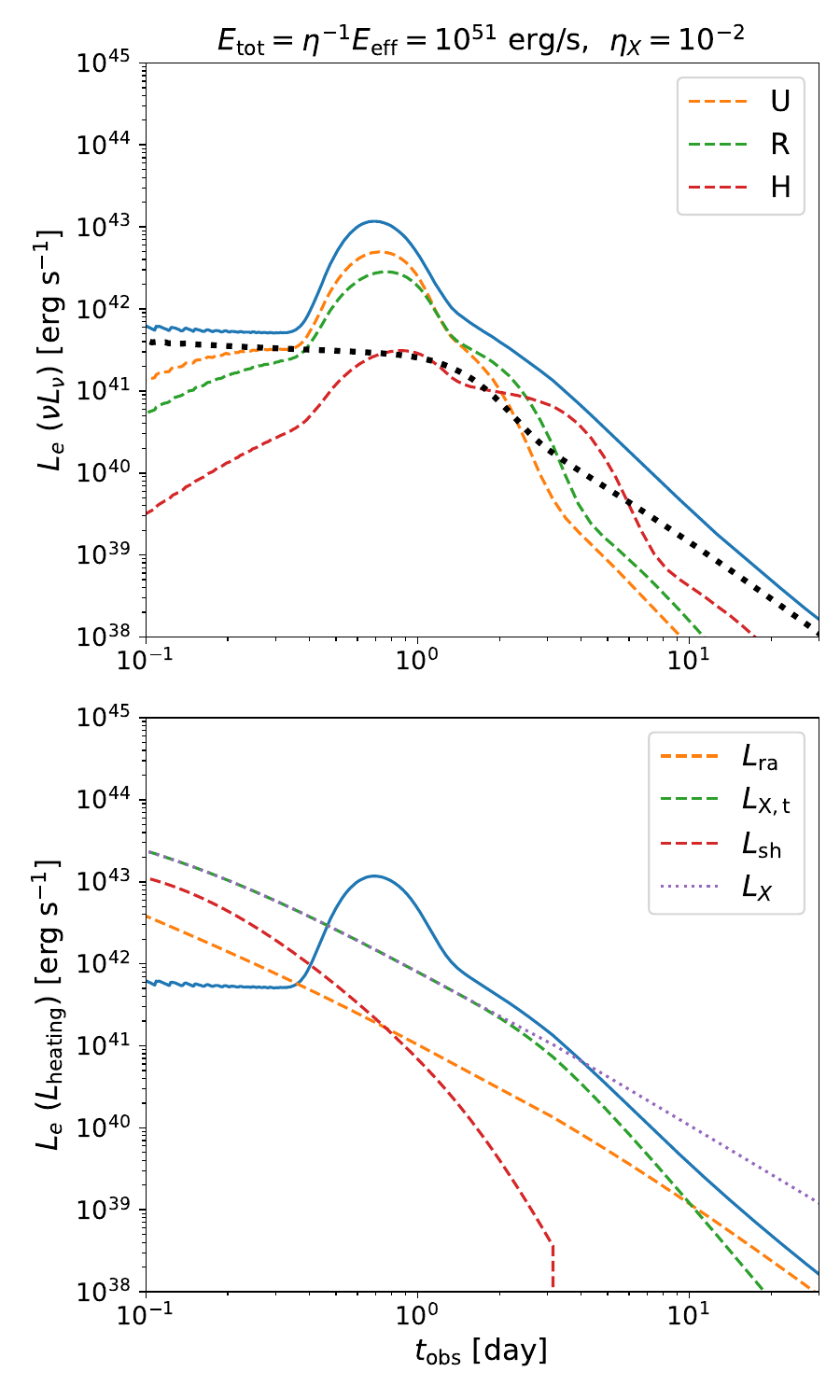}}  &
    \resizebox{55mm}{!}{\includegraphics[]{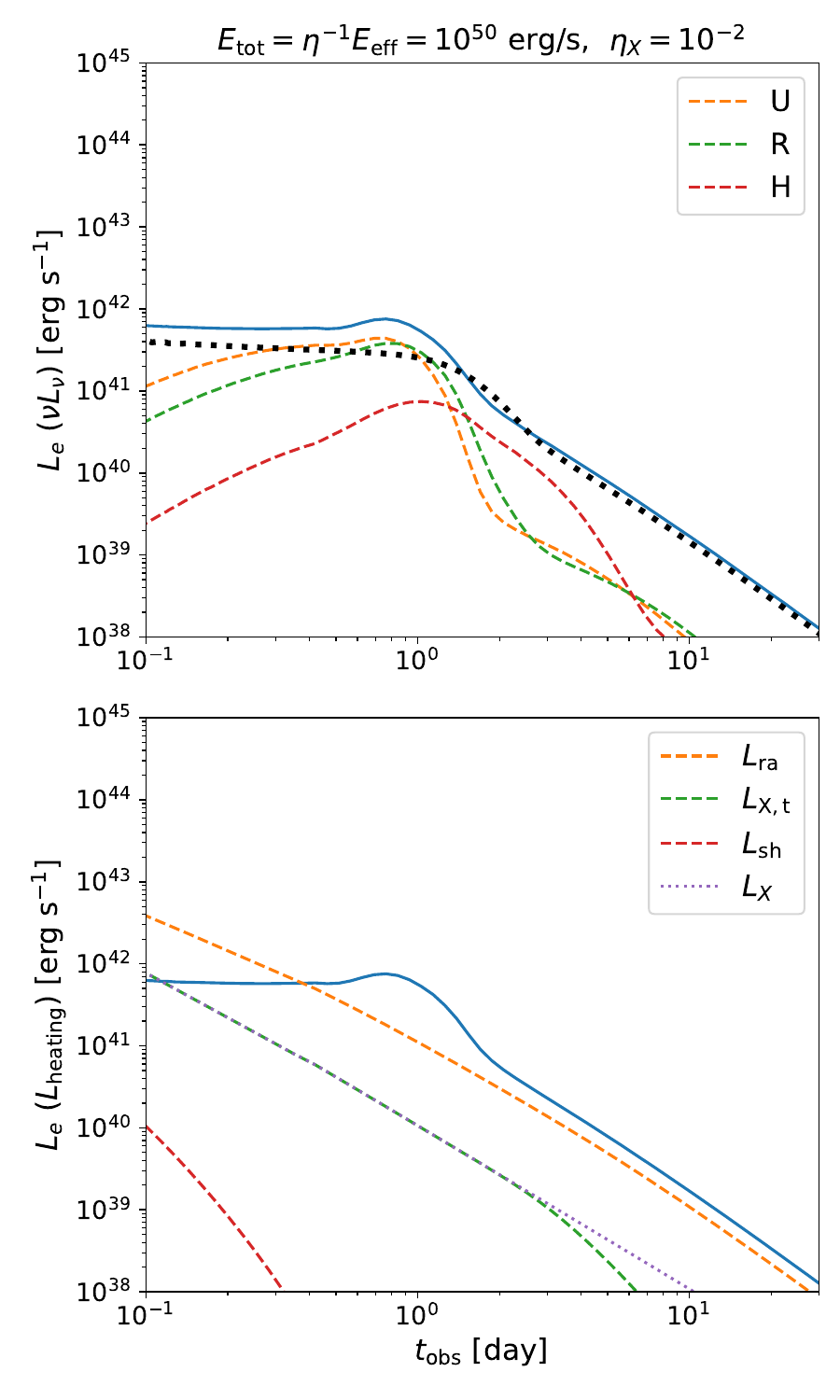}}
    \\
    \resizebox{55mm}{!}{\includegraphics[]{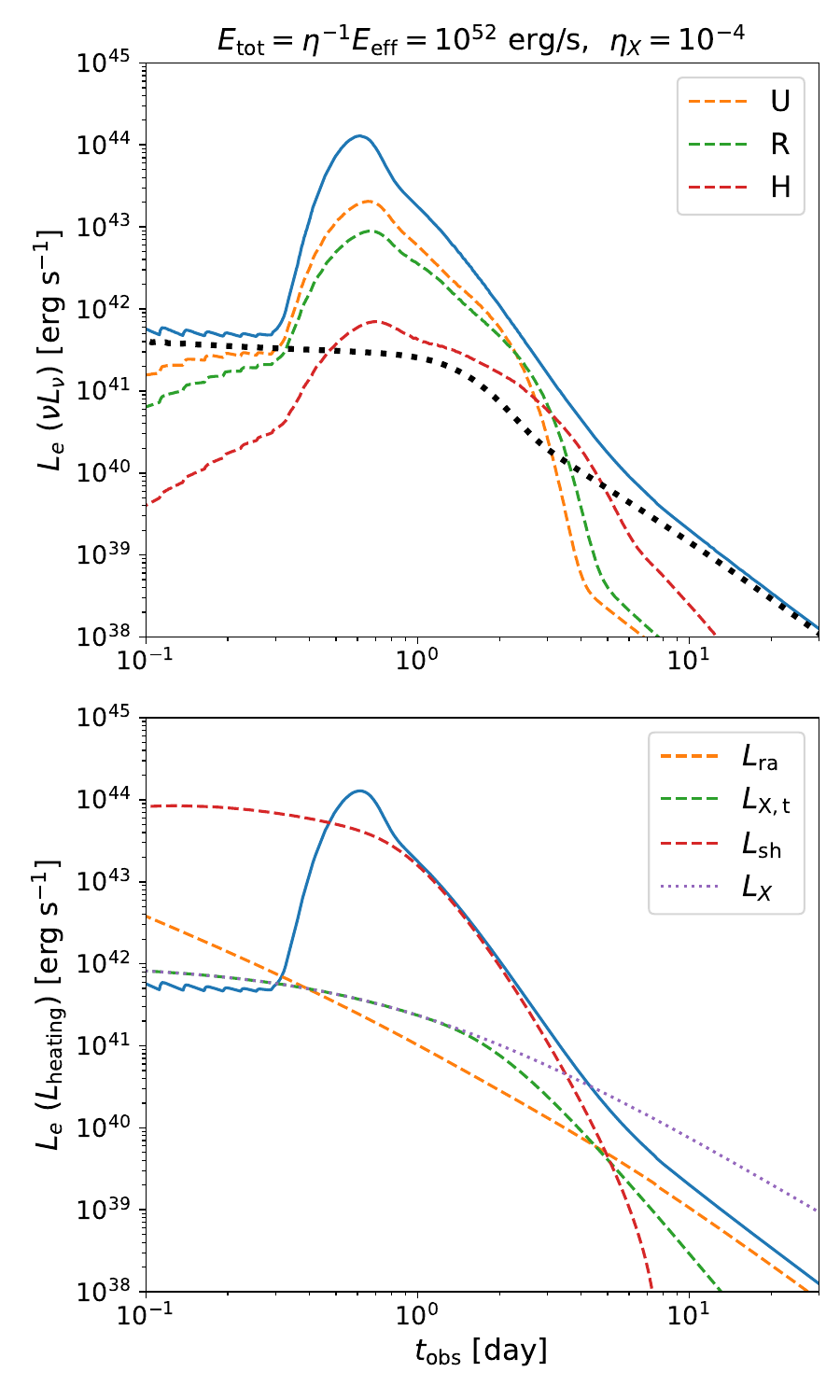}}   &  \resizebox{55mm}{!}{\includegraphics[]{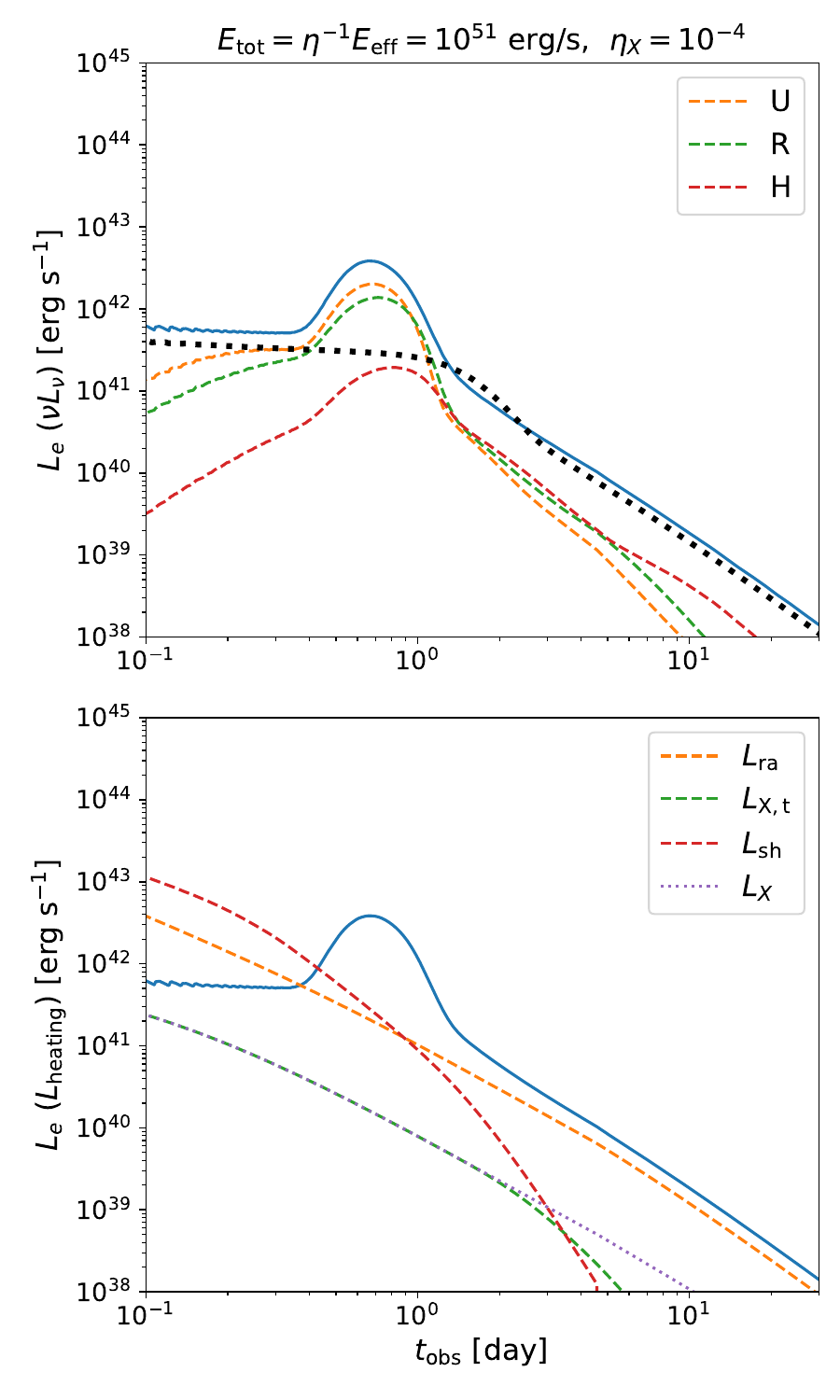}} & 
    \resizebox{55mm}{!}{\includegraphics[]{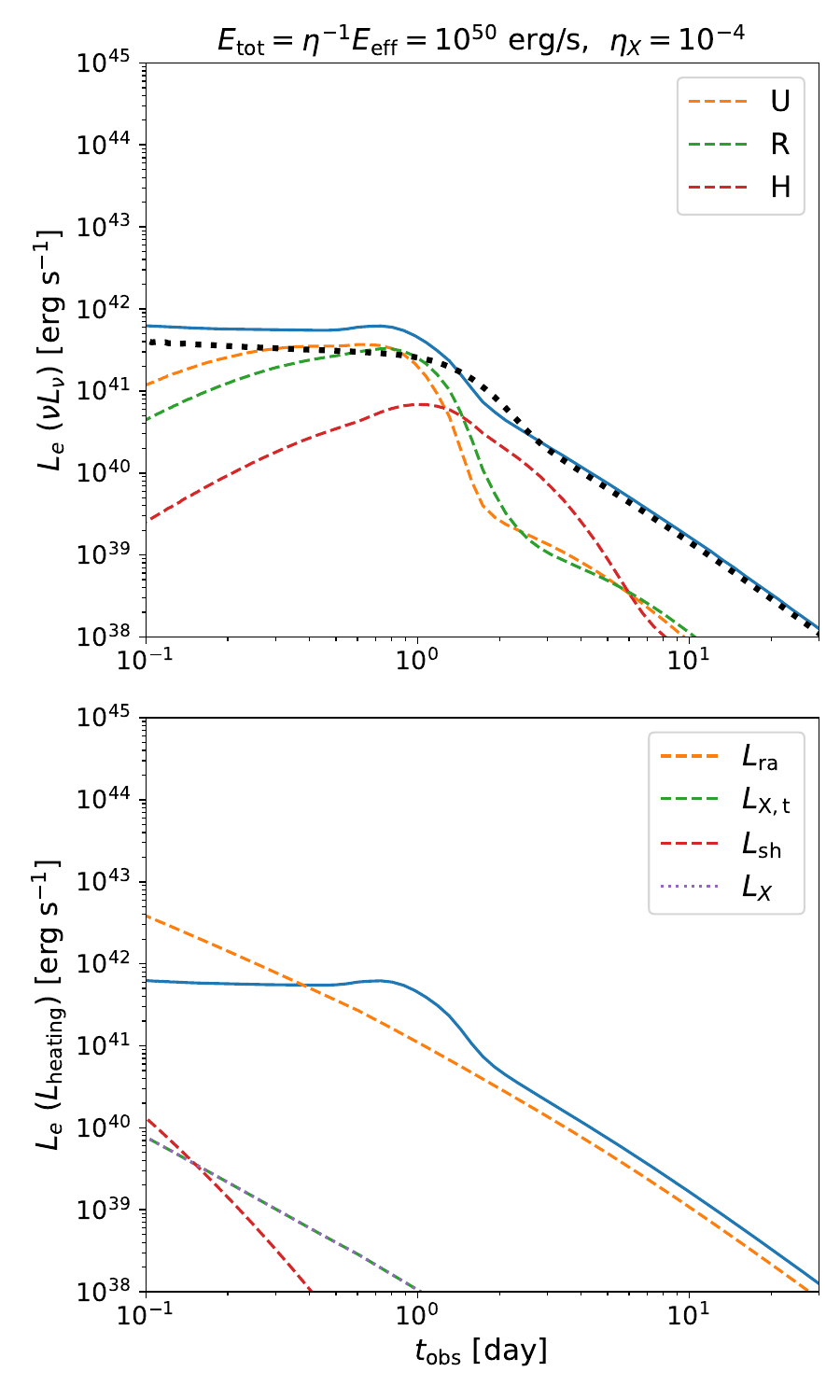}}
\end{tabular}
\caption{The light curves of engine-fed kilonovae are shown in different panels, assuming different total wind energy budgets with a fixed initial spin-down luminosity of $L_{\rm sd,0} = 10^{47}~{\rm erg/s}$. The fraction of the wind luminosity dissipated into X-rays is set to be $\eta_X = 10^{-2}$ for the upper three panels and $10^{-4}$ for the lower three panels. Across all panels, the following parameters are consistently adopted: $M_{\rm ej} = 10^{-2}M_{\odot}$, $\kappa = 1~{\rm cm^2~g^{-1}}$, $v_{\rm min} = 0.05c$, and $v_{\rm max} = 0.2c$. Each panel is further divided into two sub-panels, the upper and lower, which present light curves in distinct observational bands (``U",``R" and ``H" in the legend represent the observational bands) and the heating rates of different channels, respectively. The solid lines represent the bolometric luminosities of the engine-fed kilonovae. The black dotted lines represent the bolometric luminosity of a typical r-process kilonova without extra energy injection. The heating channels include radioactive heating ($L_{\rm ra}$), shock heating ($L_{\rm sh}$), and X-ray irradiation ($L_{X,t}$). $L_X$ represents the total X-ray luminosity released from the NS wind. }
\label{fig:engine_fed_kilonova_L47_Lx001}.
\end{figure*}

In Case I, specifically, we set $L_{\rm sd,0} = 10^{47}~{\rm erg/s}$, which requires a surface dipole magnetic field of the post-merger NS of $B_p \gtrsim 10^{14}~{\rm G}$.  As seen in Figure \ref{fig:engine_fed_kilonova_L47_Lx001}, if the spin-down timescale of the post-merger NS is shorter than the typical peak time of a r-process kilonova ($\sim 1$ day), the energy injection, especially resulting from shock heating, would primarily be efficient in the early stage \footnote{Shock heating would terminate after the shock breaks out of the ejecta.}. The injected energy would be stored inside the ejecta and ultimately released when it becomes transparent. When $E_{\rm tot} \lesssim 10^{50}~{\rm erg}$, the power provided by shock heating would not surpass that from r-process heating. The power provided by X-ray irradiation depends on the fraction, $\eta_X$, of the wind luminosity that is dissipated into X-rays. When $\eta_X \leq 10^{-2}$, the power provided by X-ray irradiation would also not exceed the r-process power. In this scenario, it might be challenging to distinguish an engine-fed kilonova from typical r-process kilonovae, particularly for events lacking early observations and with relatively low $E_{\rm tot}$.

\begin{figure*}
\centering
\begin{tabular}{ccc}
    \resizebox{60mm}{!}{\includegraphics[]{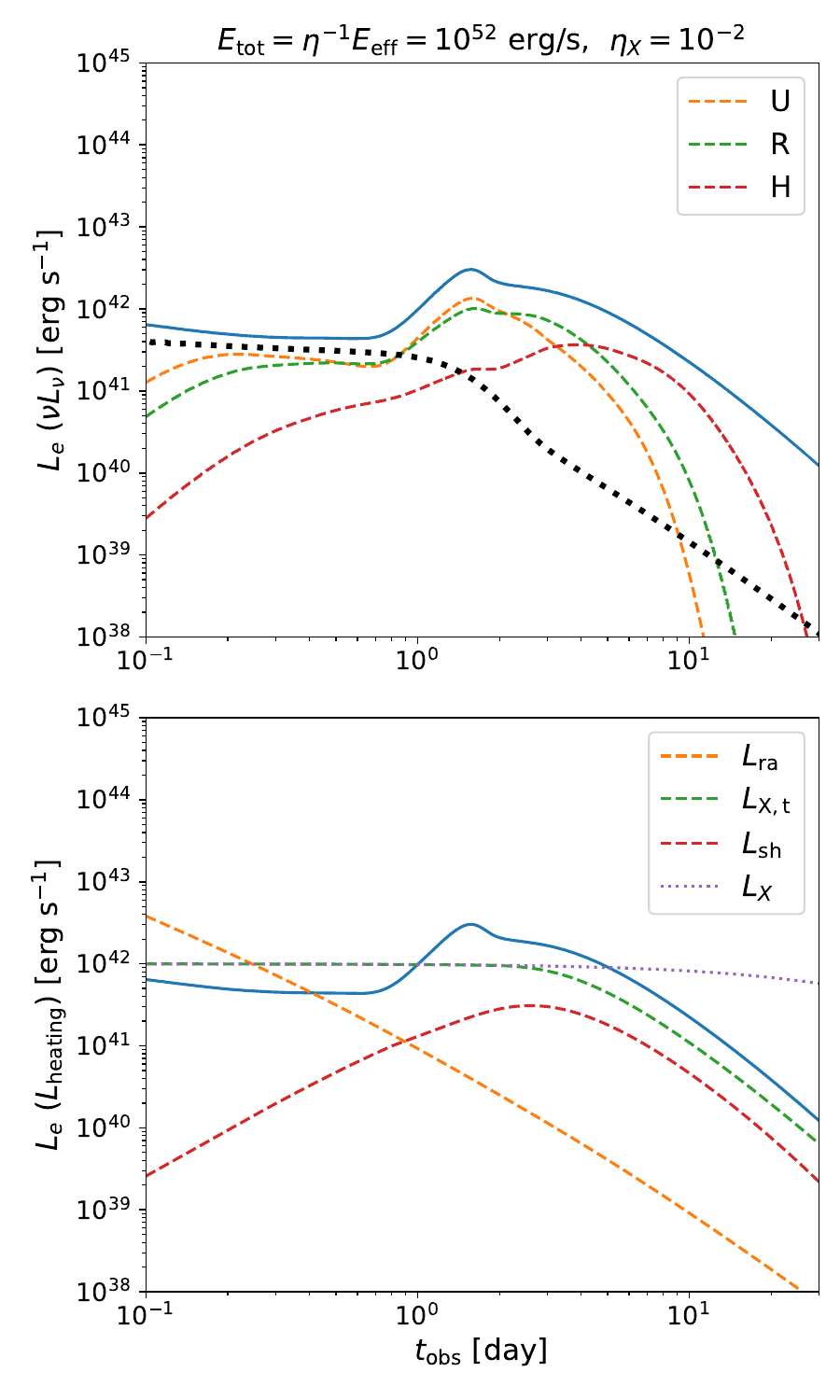}}   &  \resizebox{60mm}{!}{\includegraphics[]{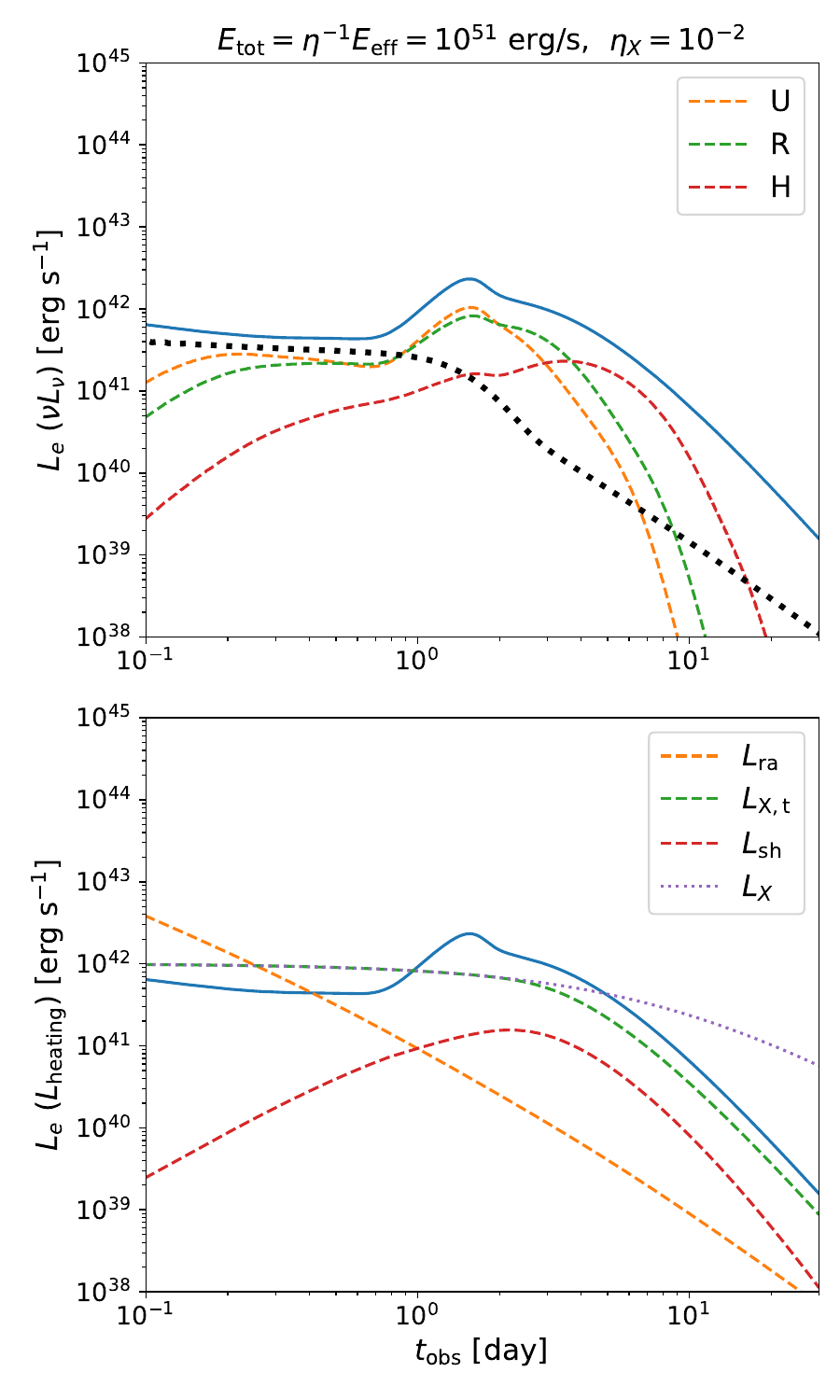}}  &
    \resizebox{60mm}{!}{\includegraphics[]{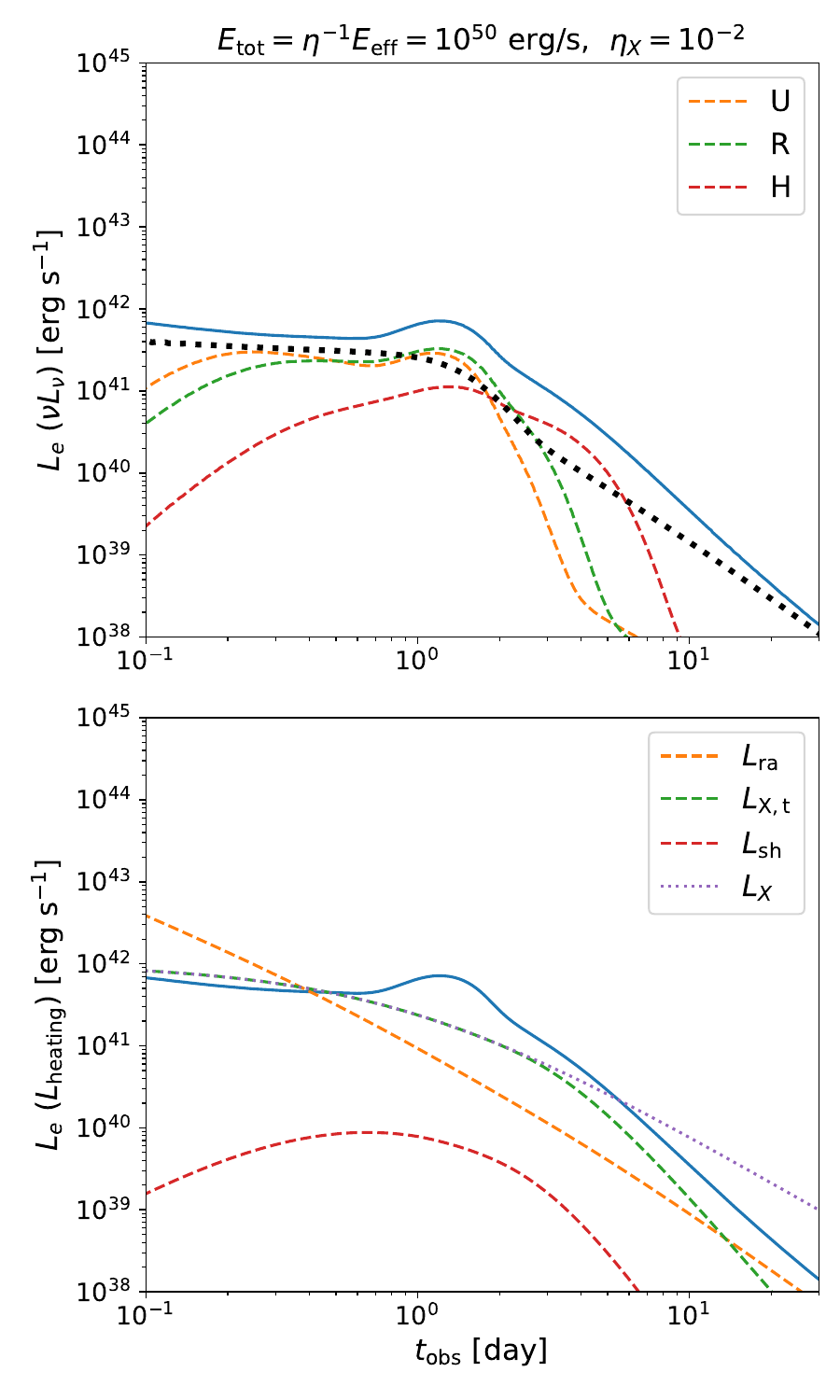}}
    \\
    \resizebox{60mm}{!}{\includegraphics[]{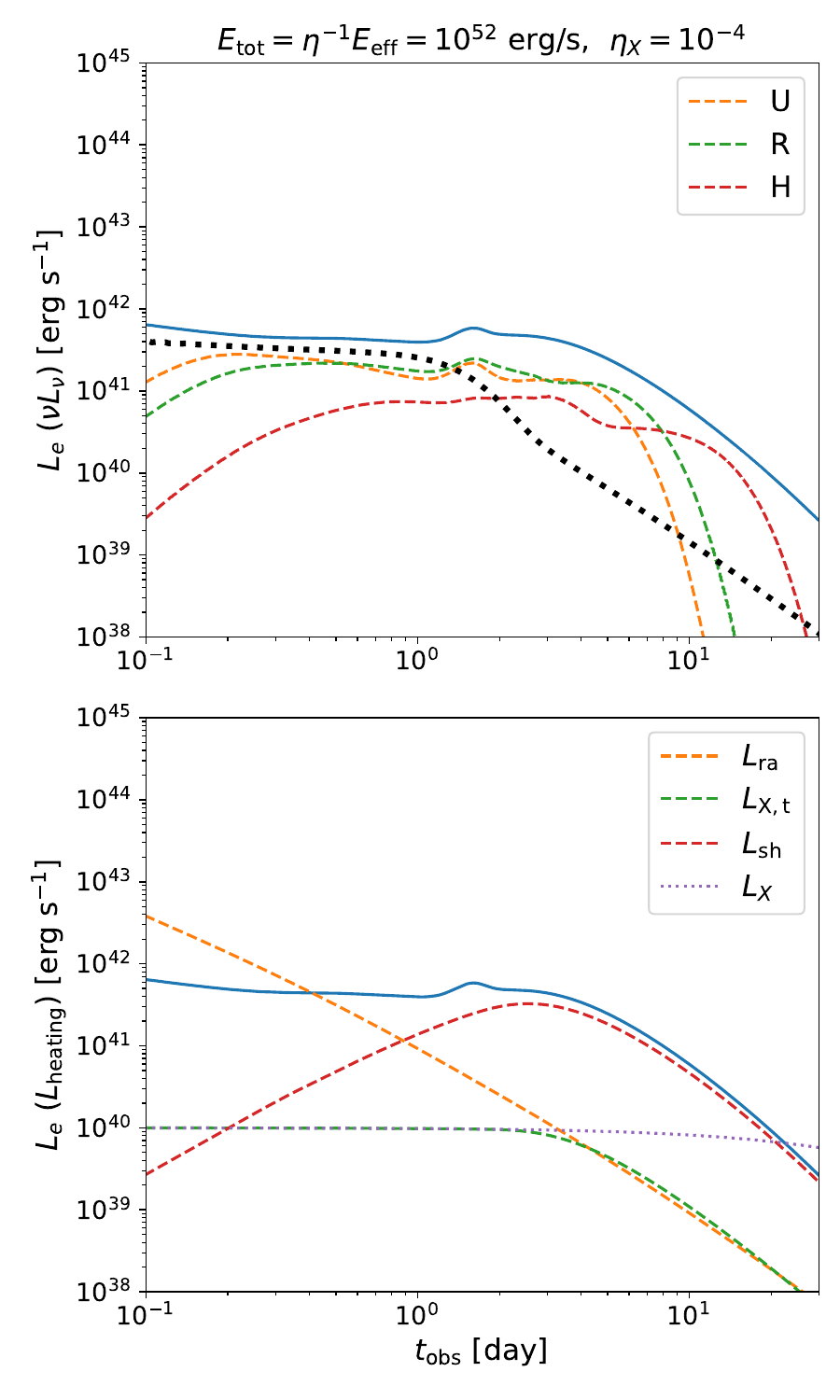}}   &  \resizebox{60mm}{!}{\includegraphics[]{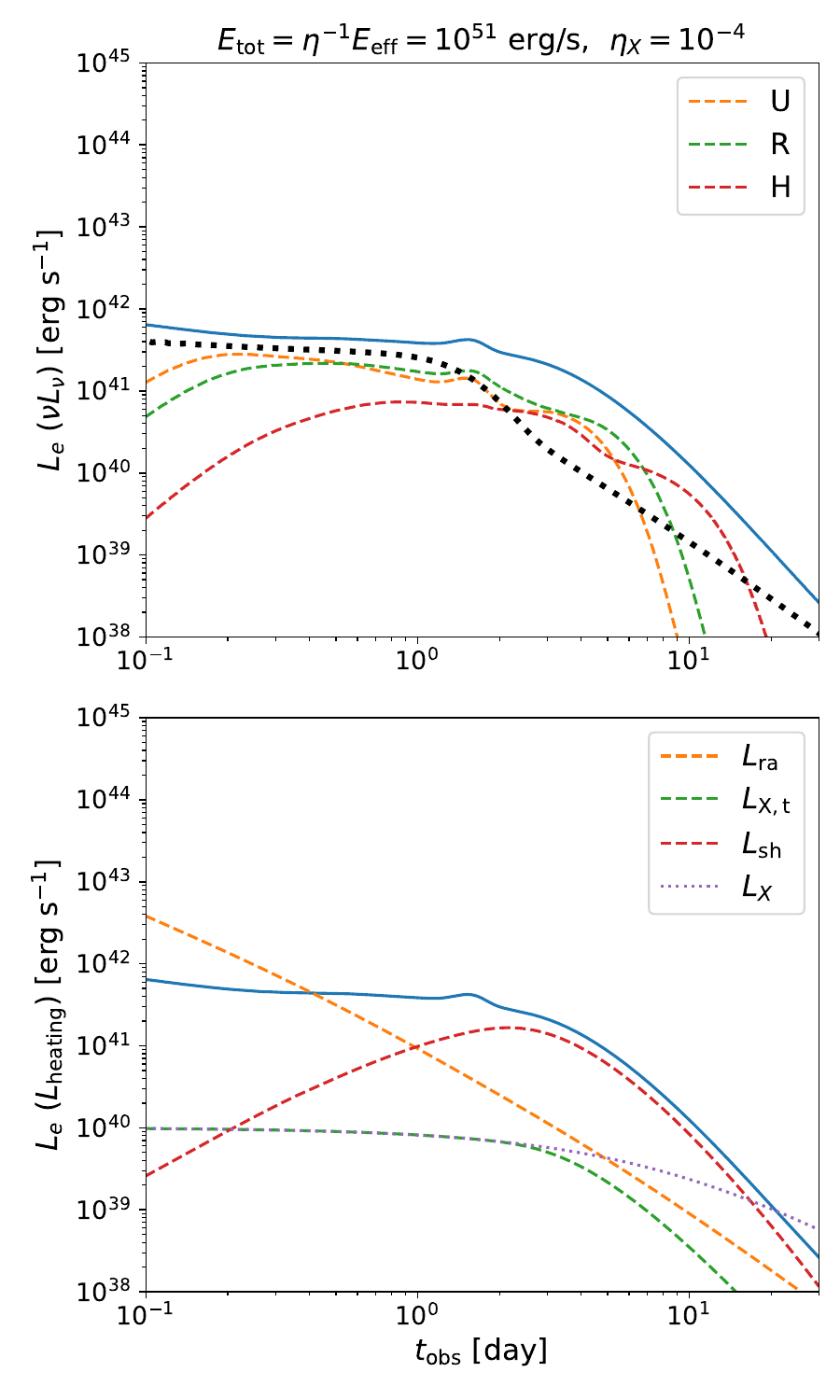}} & 
    \resizebox{60mm}{!}{\includegraphics[]{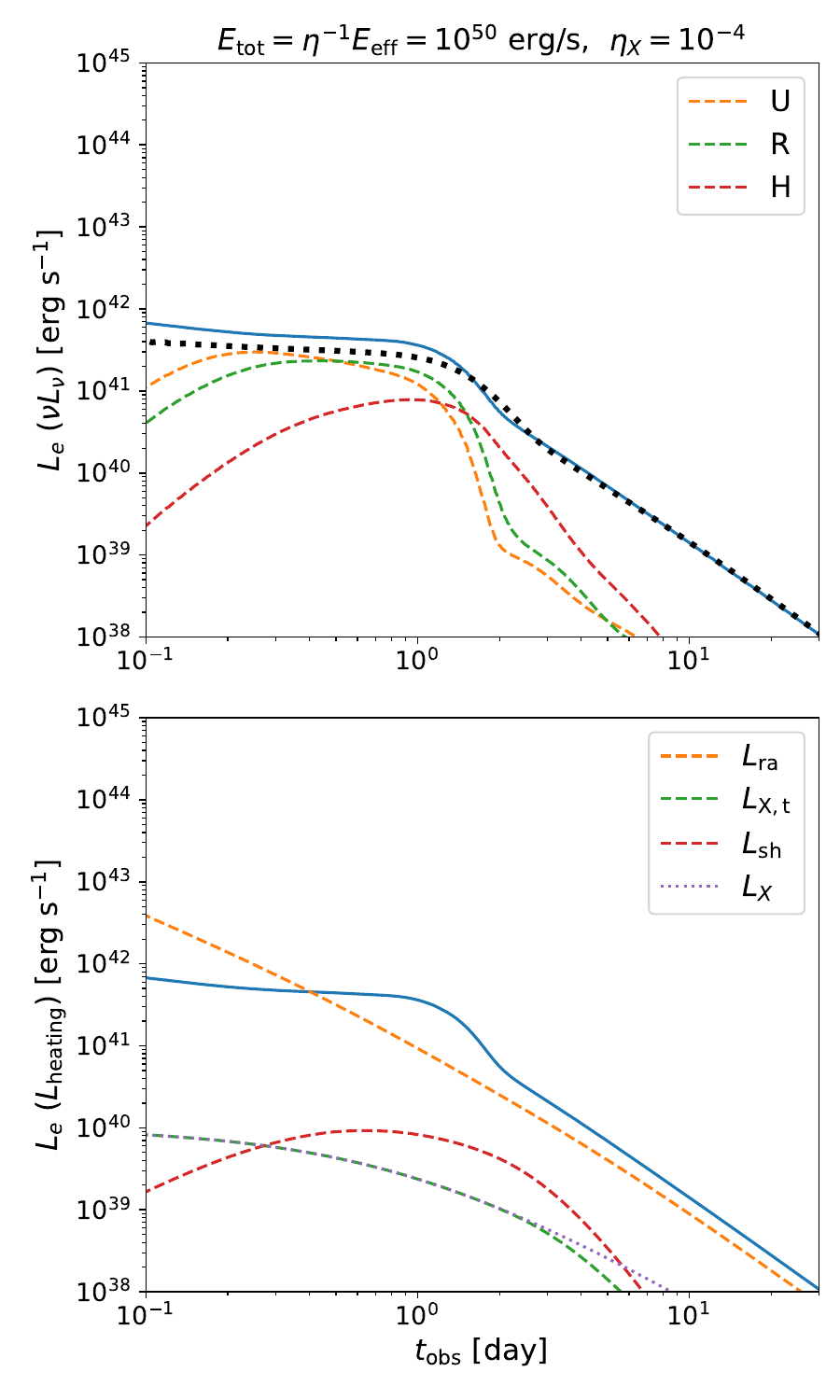}}
\end{tabular}
\caption{Similar as Figure \ref{fig:engine_fed_kilonova_L47_Lx001} but with $L_{\rm sd,0} = 10^{45}~{\rm erg/s}$ considered.}
\label{fig:engine_fed_kilonova_L45_Lx001}.
\end{figure*}

In Case II, we calculated the light curves of engine-fed kilonovae with $L_{\rm sd,0} = 10^{45}~{\rm erg/s}$, as shown in Figure \ref{fig:engine_fed_kilonova_L45_Lx001}. The energy injection from both shock heating and X-ray irradiation would be at a relatively low rate but would persist for a longer duration due to lower spin-down luminosity. Furthermore, the heating efficiency ($\xi_t$) of the FS would be reduced compared to the scenario with high spin-down luminosity, as described in our Paper I. In this case, although the peak luminosity is not as high as that in Case I, the brightening in the late stages (e.g. $> \sim 1~{\rm d}$) is obvious. Again, When $E_{\rm tot} \lesssim 10^{50}~{\rm erg}$ and $\eta_X \lesssim 10^{-2}$, the engine-fed kilonova can hardly be distinguished from the typical r-process one.

\begin{figure*}
\centering
\resizebox{85mm}{!}{\includegraphics[]{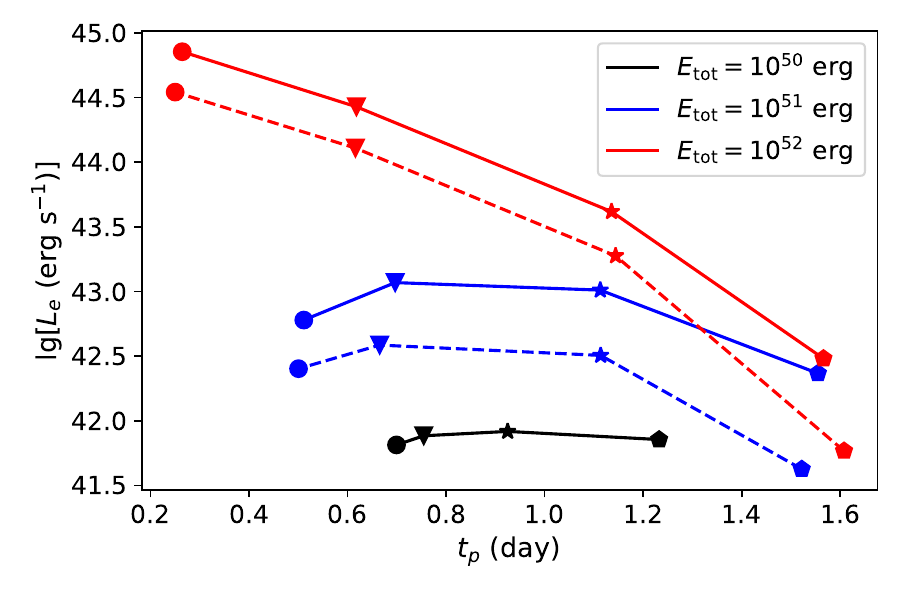}} 
\caption{Peak time and peak bolometric luminosity for engine-fed kilonovae under different parameters. The solid circles, triangles, squares and pentagons represent the cases that $L_{\rm sd,0} = 10^{48}, 10^{47}, 10^{46}, 10^{45}~{\rm erg/s}$, respectively. The solid and dashed lines represent the cases when $\eta_X = 10^{-2}$ and $10^{-4}$, respectively. Other parameters utilized are same as those in Figure \ref{fig:engine_fed_kilonova_L47_Lx001}.}
\label{fig:peaktime_peakL}.
\end{figure*}

To gain a deeper understanding of the brightness of an engine-fed kilonova signal, we present the peak time and peak luminosity of the kilonova's light curves in Figure \ref{fig:peaktime_peakL}. As can be seen, a larger total energy budget of the magnetar's wind generally leads to a higher peak bolometric luminosity. The peak time is primarily determined by the spin-down luminosity of the post-merger magnetar, where a high spin-down luminosity tends to result in an earlier peak, because the magnetar's wind might accelerate the ejecta and make it become transparent earlier. Varying the X-ray irradiation efficiency $\eta_X$ has little effect on the peak times, indicating that X-ray irradiation barely alters the ejecta dynamics. It is worth noting that a higher spin-down luminosity does not necessarily result in a higher peak luminosity of the engine-fed kilonova, particularly when the spin-down timescale of the magnetar is significantly shorter than the peak time of the light curve of the engine-fed kilonova.

We also make contour plots for the bolometric luminosities of engine-fed kilonovae under different total engine budgets and spin-down luminosities, at several different observing times:
\begin{itemize}
    \item At $t \sim 0.5$ day after the merger, the peak luminosity can reach more than $10^{44}~{\rm erg/s}$ when $E_{\rm tot} \gtrsim 4\times 10^{51}~{\rm erg}$ and $L_{\rm sd,0} \gtrsim 10^{47}~{\rm erg/s}$. When $L_{\rm sd,0} < 2 \times 10^{46}~{\rm erg/s}$, the bolometric luminosity of the engine-fed kilonova would be close to $10^{42}~{\rm erg/s}$, rougly at the same level of the typical r-process kilonova. 
    
    \item At $t \sim 1$ day after the merger, the bolometric luminosity of an engine-fed kilonova also reaches a high level ($\gtrsim 10^{43}~{\rm erg/s}$) in the case where $L_{\rm sd,0} \lesssim 10^{46}~{\rm erg/s}$. The bolometric luminosity in the case where $L_{\rm sd,0} > \sim 4 \times 10^{47}~{\rm erg/s}$ decreases to a level comparable with the typical r-process one when $\eta_X \lesssim 10^{-4}$, but remains relatively high when $\eta_X \gtrsim 10^{-2}$.
    
    \item At $t \sim 1$ week after the merger, the energy injection is still efficient for the cases with relatively low $L_{\rm sd,0}$, although it is generally fainter than that at $\sim 1$ day.     
\end{itemize}

\begin{figure*}
\centering
\begin{tabular}{cc}
  \resizebox{85mm}{!}{\includegraphics[]{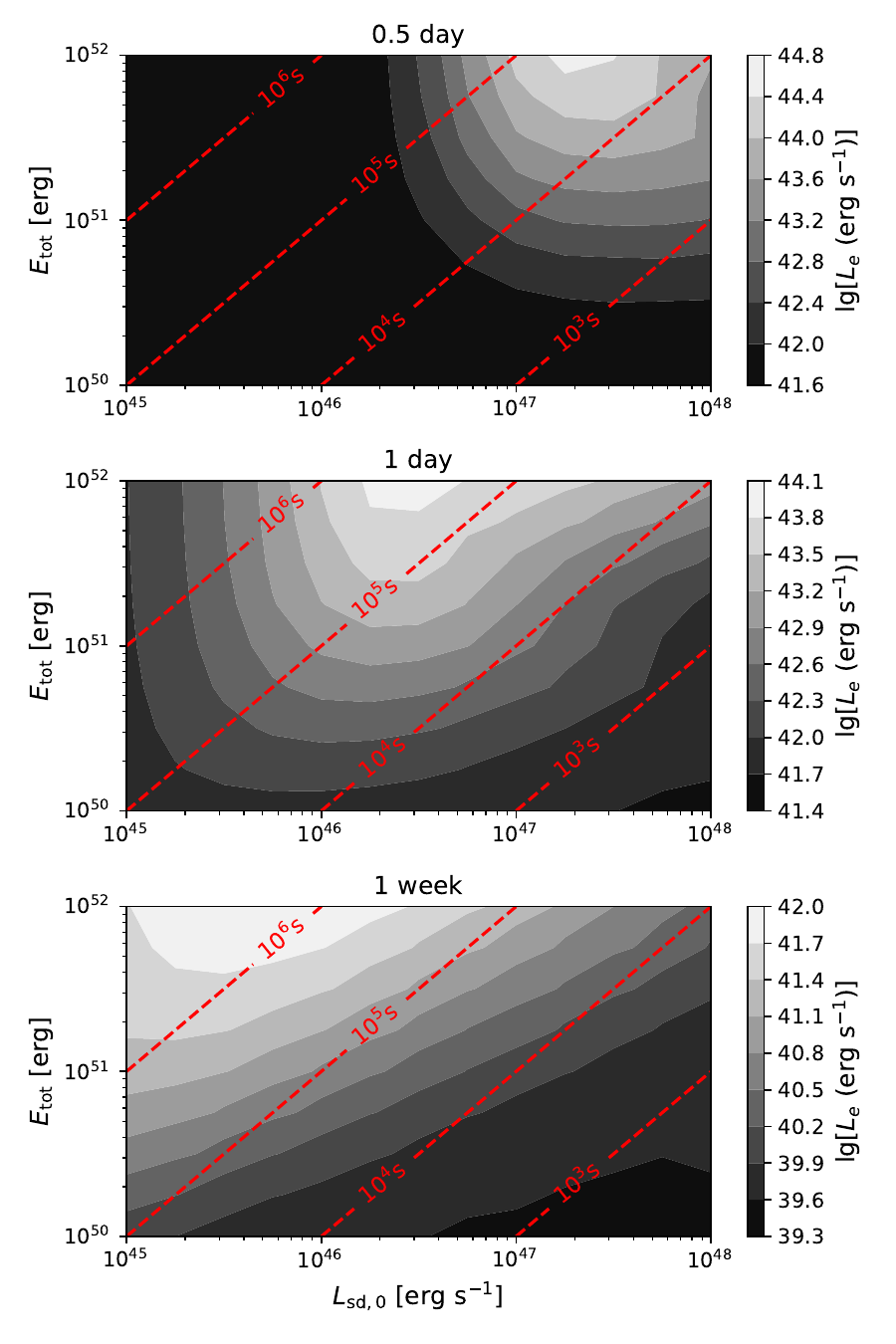}}   &  \resizebox{85mm}{!}{\includegraphics[]{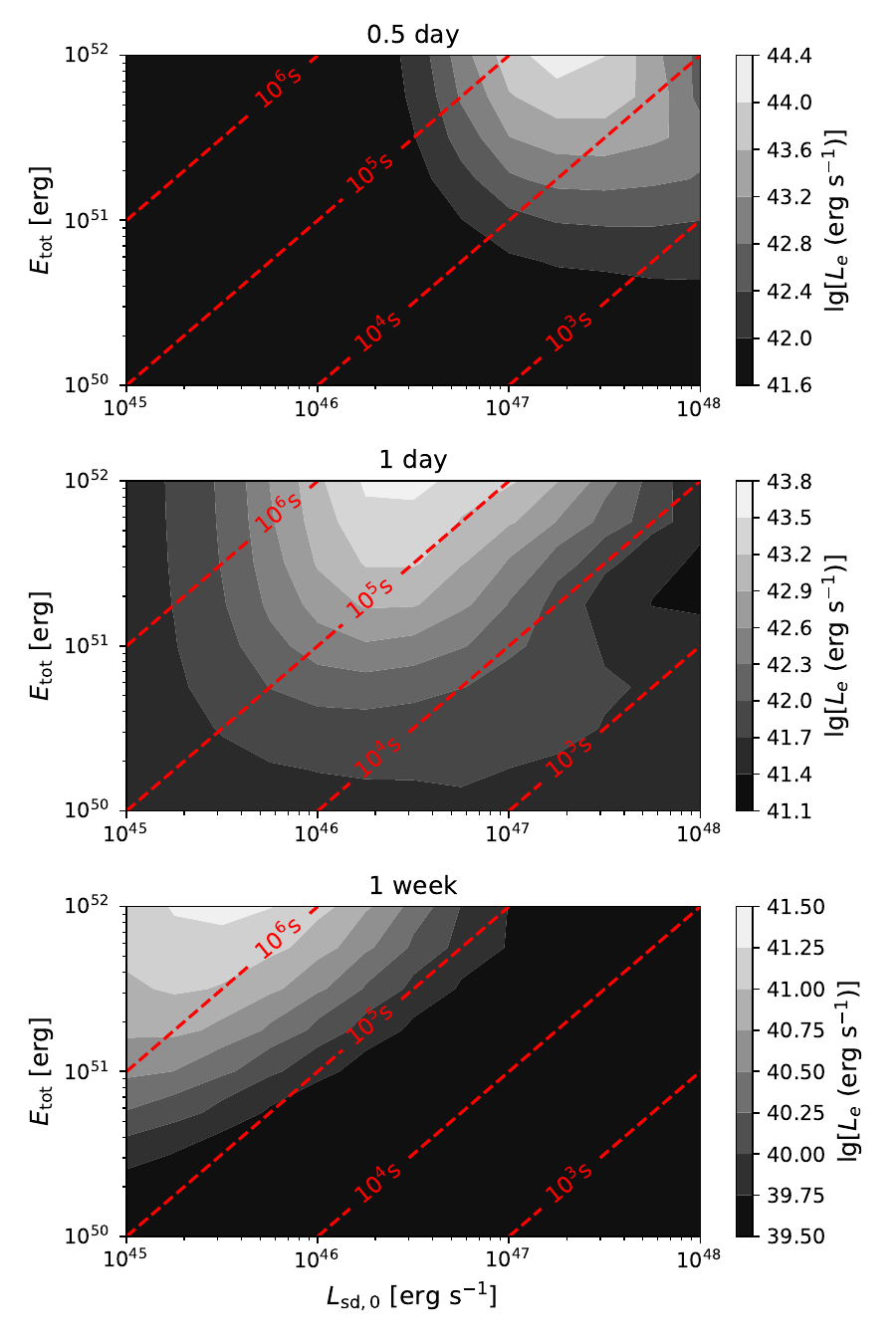}} 
\end{tabular}
\caption{The contours representing the bolometric luminosity under different total energy budgets and initial spin-down luminosities at 0.5 day, 1 day, and 1 week, respectively, are shown with a grey color map. The contours for the spin-down timescale ($t_{\rm sd}$) are shown with dashed lines. The left and right panels represent the cases when $\eta_X = 10^{-2}$ and $10^{-4}$, respectively. Other parameters utilized are same as those in Figure \ref{fig:engine_fed_kilonova_L47_Lx001}.}
\label{fig:L_contour}.
\end{figure*}

\subsection{Spectral energy distribution}
\label{sec:SED}
\begin{figure*}
\centering
\begin{tabular}{cc}
  \resizebox{85mm}{!}{\includegraphics[]{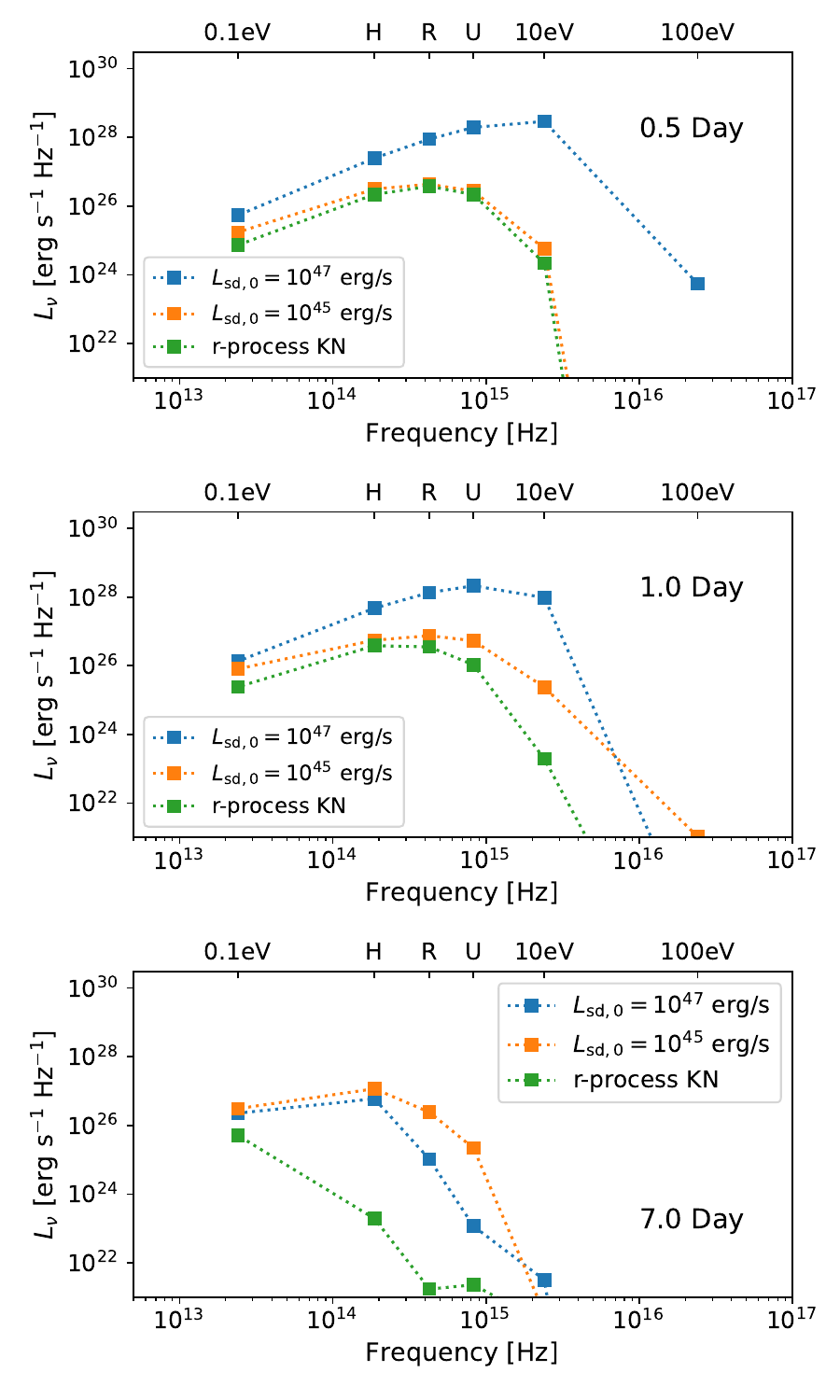}}   &  \resizebox{85mm}{!}{\includegraphics[]{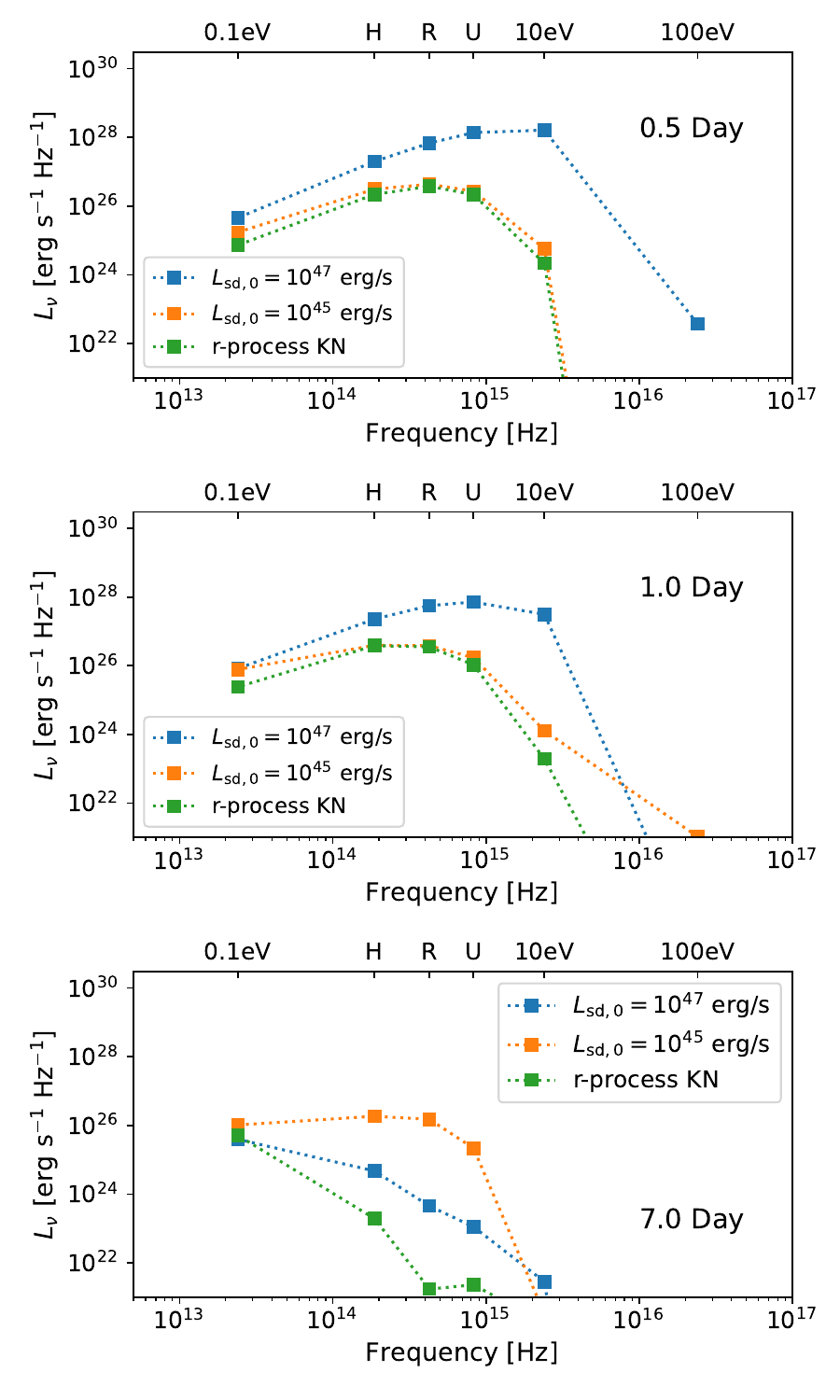}} 
\end{tabular}
\caption{The Spectral energy distributions of (engine-fed) kilonovae at different times. The left and right panels represent the cases when $\eta_X = 10^{-2}$ and $10^{-4}$, respectively. The total wind energy budget is adopted to be $E_{\rm tot} = 10^{52}~{\rm erg}$. Other parameters utilized are the same as those in Figure \ref{fig:engine_fed_kilonova_L47_Lx001}.}
\label{fig:SED}.
\end{figure*}

Intuitively, due to the energy injection, the temperature inside the merger ejecta would increase, resulting in a bluer engine-fed kilonova, as shown in Figure \ref{fig:SED}.

With a relatively high spin-down luminosity ($L_{\rm sd,0} \sim 10^{47}~{\rm erg/s}$), a UV bump appears earlier than the typical peak time of an r-process kilonova, and then gradually fades over time. Meanwhile, the engine-fed kilonova becomes redder, with the peak frequency shifting to the optical band at approximately $1$ day and the infrared band at approximately $1$ week. At a relatively late stage ($\gtrsim 1$ day), although the bolometric luminosity of the engine-fed kilonova becomes comparable to the typical r-process one, as demonstrated in Figure \ref{fig:engine_fed_kilonova_L47_Lx001}, the SED should be bluer. This is attributed to the compression of the ejecta volume by the FS-RS system, resulting in a higher temperature with even the same internal energy.

With a relatively low spin-down luminosity of $L_{\rm sd,0} \sim 10^{45}~{\rm erg/s}$, there is almost no difference between the SED of the engine-fed kilonova and the typical r-process kilonova before they both become effectively transparent at approximately $1$ day. Subsequently, due to the continuous energy injection from the magnetar wind, the cooling process of the engine-fed kilonova is slower, resulting in an overall temperature that remains higher than a typical r-process kilonova. Consequently, in the relatively late stage, the engine-fed kilonova appears bluer than the r-process one.

\subsection{Shock breakout}
\label{sec:shock_breakout}
\begin{figure*}
\centering
\begin{tabular}{cc}
\resizebox{85mm}{!}{\includegraphics[]{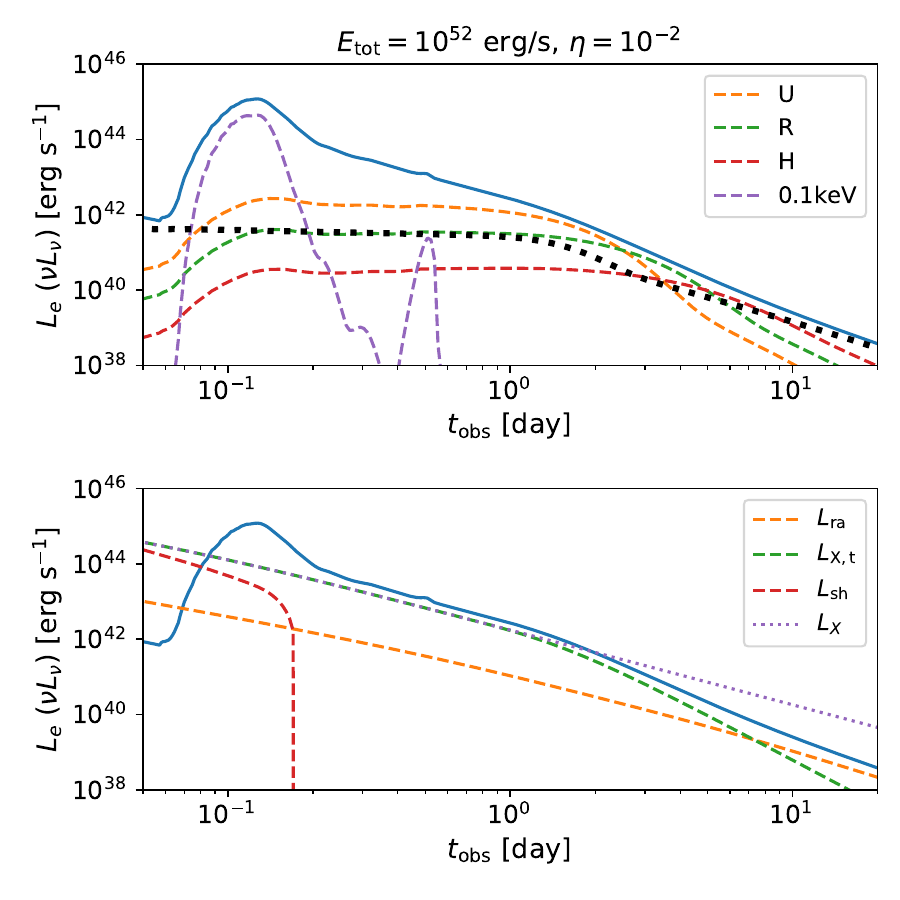}} & \resizebox{85mm}{!}{\includegraphics[]{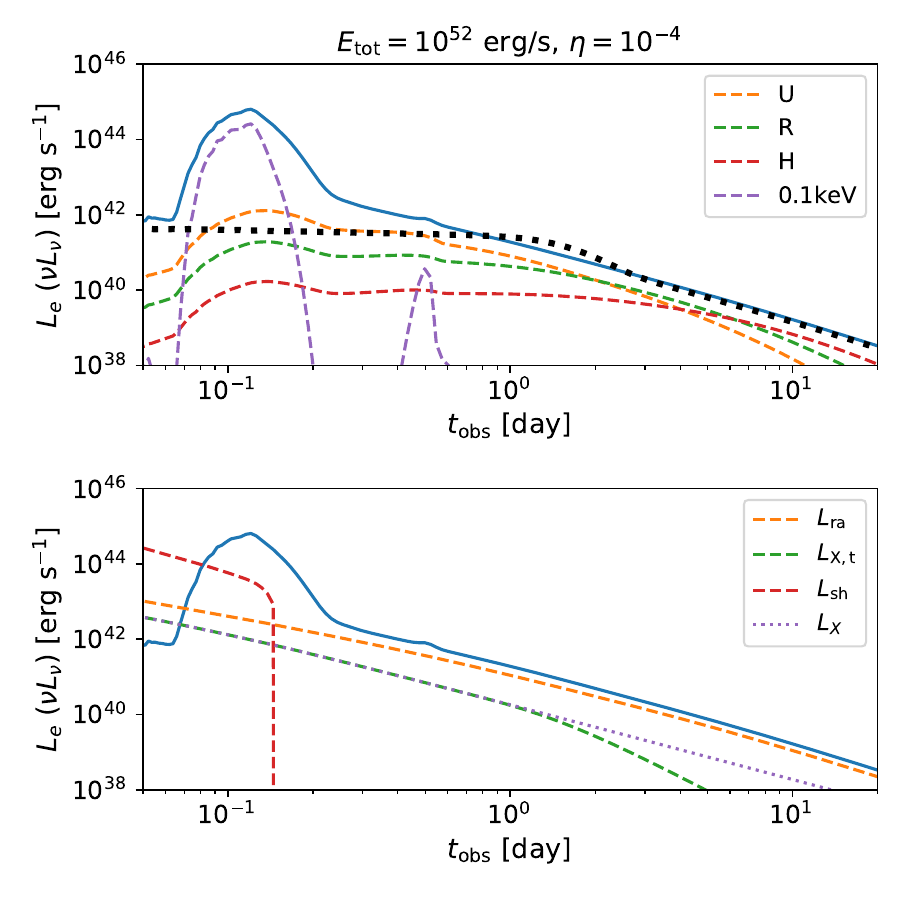}} 
\end{tabular}
\caption{Similar as Figure \ref{fig:engine_fed_kilonova_L47_Lx001} but with $L_{\rm sd,0} = 5 \times 10^{48}~{\rm erg/s}$ considered. A shock breakout signal at $0.1{\rm keV}$ is predicted. The left and right panels represent the cases when $\eta_X = 10^{-2}$ and $10^{-4}$, respectively.}
\label{fig:shock_breakout}.
\end{figure*}

When both the spin-down luminosity and the total energy budget of the magnetar wind are sufficiently large, the FS will quickly break out from the ejecta, producing a prominent signal in the soft X-ray band. The shock breakout signal associated with an engine-fed kilonova has been investigated by \cite{lisz2016}. In their study, the internal energy provided by shock heating was assumed to accumulate right behind the FS and be released shortly after the shock escapes the ejecta. Consequently, the shock breakout signal in the temporal domain was presumed to be exceedingly brief. In contrast, this work assumes that shock heating occurs where the material is being swept, and models the heat propagation (photon diffusion) within the radiation-dominated ejecta. Consequently, both the ascending and descending phases may not be as sharp as those described in \cite{lisz2016}. Nevertheless, the shock breakout signal is anticipated to last for less than $1$ day. It remains less apparent in other bands, such as UV and optical, where the light curves resemble a plateau. This is because the shock breakout signal is closely adjacent to the main peak of the kilonova signal, which occurs approximately at $1$ day after the merger.

\section{Conclusion and discussion}
\label{sec:conclusion}
If a long-lived NS (instead of a promptly collapsing black hole or a short-lived hyper-massive NS) is produced after a BNS merger, the NS wind would inject extra energy into the merger ejecta and power an engine-fed kilonova. We have developed a code to calculate the multi-band light curves of an engine-fed kilonova. The dynamics of the wind-ejecta interaction is based on the mechanical model of magnetized relativistic blastwaves, which has been described in detail in the first paper of this series \citep{ai2022}. Taking into account the heating effects due to both the shock excited in the ejecta and the X-ray irradiation from somewhere behind the ejecta, we calculated the photon diffusion inside the ejecta and the thermal radiation emitted from it.

The light curves and spectral energy distributions of engine-fed kilonovae demonstrate a diverse behavior based on the spin-down luminosity of the post-merger NS and the total energy injection budget. This diversity can be categorized into two cases:
\begin{itemize}
    \item In the case where the spin-down luminosity is relatively high, for instance, $L_{\rm sd,0}\sim 10^{47}~{\rm erg/ s}$, a prominent early blue bump is anticipated. If the spin-down timescale is significantly shorter than the peak time of the engine-fed kilonova, the bolometric luminosity would decrease to a level comparable to that of the typical r-process kilonova approximately $1$ day after the merger. The SED of the engine-fed kilonova would appear bluer than the typical r-process kilonova, in both the early and late stages.

    \item In the case where the spin-down luminosity is relatively low, for instance, $L_{\rm sd,0}\sim 10^{45}~{\rm erg/s}$, the early blue bump would not be evident. Instead, the prolonged shock propagation and X-ray irradiation would lead to a substantial brightening of the kilonova in the relatively late stage. The SED of the engine-fed kilonova would resemble that of the typical r-process kilonova before approximately $1$ day, but it would be significantly bluer than the typical r-process kilonova in the late stages, due to the continuous energy injection.
\end{itemize}
Additionally, a shock breakout signal is anticipated when both the spin-down luminosity and the total energy budget of the NS (magnetar) wind are sufficiently large. This signal is predicted to occur in the soft X-ray band and emerge approximately $0.1$ day after the merger, lasting for less than $1$ day. It is noteworthy that, within a large parameter space, especially when the total energy injection budget is less than approximately $10^{50}~{\rm erg}$, it can be challenging to distinguish the engine-fed kilonova from the typical r-process ones.

If the X-ray plateau in the afterglow light curve of GRB 230307A is indeed powered by a post-merger magnetar, the break time of the plateau ($\sim 100$ s) should be equal to the magnetar's spin-down timescale. Based on the redshift obtained from the most likely host galaxy, the luminosity of the X-ray plateau is expected to be $\sim 10^{48}~{\rm erg/s}$ \citep{sun2023}. This luminosity should originate from the polar region of the spin axis. Assuming a beaming factor of the NS wind due to rapid spin of $\eta \sim 0.1$ and a dissipation efficiency of the magnetic field of $\eta_X = 10^{-2}$ ($10^{-1}$), the spin-down luminosity of the magnetar is presumed to be $L_{\rm sd,0}\sim 10^{49}~{\rm erg/s}$ ($10^{48}~{\rm erg/s}$). Consequently, the total energy budget for the NS wind is $E_{\rm tot} = 10^{51}~{\rm erg}$ ($10^{50}~{\rm erg}$). As shown in Figure \ref{fig:L_contour}, since the ultraviolet/optical/infrared observations began approximately $1$ day after the merger, the early blue bump may have been missed. Therefore, the non-recognition of the feature of an engine-fed kilonova cannot rule out the possibility that a magnetar was produced at this moment. Whether our engine-fed kilonova model can explain the observation of GRB 230307A's kilonova remains to be determined through fitting together with the afterglow data. In the literature, there also exist models that suggest neutron star-white dwarf mergers as potential explanations for the occurrence of long gamma-ray bursts (GRBs) associated with kilonova-like signal \citep{yang2022,zhong2023,wangxy2024}. In this scenario, the ``kilonova" may be attributed to the decay of $^{56}Ni$ combined with energy injection from a post-merger neutron star \citep{zhong2023,wangxy2024}. In principle, our model can be adapted to accommodate this explanation by simply replacing the empirical formula for r-process heating with one based on the decay of $^{56}Ni$.

Intuitively, the decay of the early bump in the case with high spin-down luminosity may mimic the ``blue'' kilonova component detected early on and the later ``red'' component can be interpreted by the late predicted lightcurve, in a GW170817-like event. Similar to that for GRB 230307A, it is necessary to consider that the early peak has been missed (peaks at $t_{\rm obs} < 0.5$ day), resulting in the requirement for a relatively large spin-down luminosity ($L_{\rm sd,0} \gtrsim 10^{47}~{\rm erg/s}$), as demonstrated in Figure \ref{fig:peaktime_peakL}. If this is the case, the spin-down timescale for the central magnetar should be shorter than $\sim 10^5~{\rm s}$, leading to an energy injection rate that follows $L \propto t^{-2}$ at later times. However, observations of the bolometric luminosity evolution for the kilonova associated with GW170817 at late times approximately follow $L \propto t^{-1}$ \citep{drout2017}, which roughly aligns with the evolution of the radioactive heating rate. Therefore, in this scenario, the late ``red'' component is likely powered by radioactive heating. From another perspective, the late ``red'' component may be explained by continuous energy injection from a magnetar with either moderate or low spin-down luminosity. However, for the moderate spin-down luminosity case ($L_{\rm sd,0} \sim 10^{45} - 10^{46} {\rm erg/s}$), a bump at $t\sim 1~{\rm s}$ is expected, violating early observations. For the low spin-down luminosity case, the early bump does not appear, so that the ``blue'' component is dominated by radioactive power. Generally speaking, either the ``blue'' or ``red'' component should be dominated by radioactive power. Using our comprehensive engine-fed kilonova model, while the constraints on the dipole magnetic field on the surface of the post-merger NS in \cite{ai18} could be somewhat relaxed, whether a single-component ejecta with additional energy injection can adequately explain the GW170817 kilonova, again, remains to be determined through further fitting processes, together with GRB afterglow data.

There is another model in the literature that predicts an early blue bump on the kilonova light curve, resulting from the ejecta being shocked by the GRB jet \citep{wu2022,hamidani2024a,hamidani2024b}, without the introduction of a post-merger magnetar. The signal powered by the GRB jet has a strong angular dependence, which is concentrated in the nearly polar direction. By contrast, our spin-down luminosity-powered, engine-fed kilonova can be observed in a wider range of observing angles.

The existence of a post-merger magnetar not only alters the dynamics and temperature of ejecta, but could also affect its opacity. Initially, neutrino irradiation emitted by the newly formed NS increases the electron fraction of the ejecta, effectively impeding the formation of heavy r-process elements \citep{metzgerpiro2014,lippuner2017}. Subsequently, persistent X-ray irradiation modifies the ionization state of heavy elements, altering the bound-free absorption of kilonova photons \citep{metzgerpiro2014,wang2023}. Consequently, assuming a constant opacity is an oversimplification. Future works involving detailed modeling of the radiation transfer in the ejecta, taking into account photon ionization, are necessary to obtain more precise light curves, spectra, and polarization information.

\section*{Acknowledgements}
This work is supported by the China Postdoctoral Science Foundation (2023M732713), the National Natural Science Foundation of China (Projects 12021003), and the National SKA Program of China (2022SKA0130100).

\section*{Data Availability}
The Python code developed to solve the theoretical problem is public \citep{ai2024code}.


\label{lastpage}
\end{document}